\documentclass[a4paper,11pt]{article}

\pdfoutput=1 

\usepackage{jheppub} 

\usepackage[T1]{fontenc} 

\usepackage{physics}
\usepackage[dvipsnames, svgnames,  x11names ]{xcolor}
\usepackage{tikz}
\usetikzlibrary{arrows,snakes,backgrounds}
\newcommand{\RNum}[1]{\uppercase\expandafter{\romannumeral #1\relax}}

\title{\boldmath Entanglement Wedge Reconstruction Beyond the  Large $N$ Limit via the Twirled Petz Map}

\author{

}
\author[a]{Arash Alipour Shahmiri,}
\author[a]{Maryam Sharifian,}
\author[a,b]{and Niloofar Vardian}

\affiliation[a]{Department of Physics, Sharif University of Technology,\\
P.O. Box 11155-9161, Tehran, Iran}
\affiliation[b]{Research Center for High Energy Physics, Department of Physics, Sharif University of Technology, \\ P.O. Box 11155-9161, Tehran, Iran}

\emailAdd{arash.alipour317@sharif.edu}
\emailAdd{maryam.sharifian@sharif.edu}
\emailAdd{niloofarvardian@gmail.com}

\abstract{  
Entanglement wedge reconstruction provides the sharpest formulation of bulk locality in AdS/CFT, identifying the entanglement wedge as the largest bulk region reconstructible from a boundary subregion. While this statement is well understood at leading order in the large 
$N$ expansion—through both modular flow and Petz map reconstruction—its extension beyond the semiclassical regime has remained unclear. In this work, we show that quantum error correction provides the natural framework for addressing this problem. At leading order, we reproduce the known leading-order reconstruction formulas, and by employing the twirled Petz map, we obtain a systematic and explicit method for incorporating subleading 
$1/N$ corrections. Beyond the leading approximation, we have obtained a bi-local correction controlled by a subleading Petz kernel. Our results establish a controlled approach to bulk reconstruction beyond large 
$N$ and sharpening the role of quantum error correction in holography. This framework provides a controlled path toward understanding bulk reconstruction in fully quantum gravitational settings.
}

\begin{document} 

\makeatletter
\gdef\@fpheader{}
\makeatother

\maketitle
\flushbottom

\section{Introduction}

A central problem in quantum gravity and holography is to understand how local bulk observables emerge from a nonlocal boundary quantum field theory. In the AdS/CFT correspondence, while the extrapolate dictionary provides a relation between boundary operators and bulk fields near the asymptotic boundary, it does not explain how operators deep in the interior of spacetime are encoded in the boundary theory. The resolution of this problem is known as bulk reconstruction, and it lies at the heart of questions concerning locality, causality, and the quantum structure of spacetime.

A particularly sharp formulation of bulk reconstruction arises through subregion duality. Given a spatial region 
$A$ of the boundary CFT, one may ask which bulk operators can be reconstructed using only operators supported in 
$A$. Early considerations based on causal structure suggested that reconstruction should be limited to the causal wedge of 
$ A$. However, it is now understood that the correct bulk region associated with a boundary subregion is generically larger \cite{Headrick:2014cta, Wall:2012uf} and is instead given by the entanglement wedge 
$ a= \mathcal{E}_A$ \cite{dong2016reconstruction}. The entanglement wedge is defined as the bulk domain of dependence of a spacelike surface bounded by 
$A$ and its corresponding quantum extremal surface 
$\chi_A$. Importantly, it is believed that 
$\mathcal{E}_A$ is the largest bulk region whose operators can be reconstructed from boundary data in 
$A$, making entanglement wedge reconstruction (EWR) the strongest and most complete realization of subregion duality consistent with holography.
EWR asserts that for any operator 
$\phi(X)$ with 
$X \in a = \mathcal{E}_A$, there exists a boundary representation 
$\Phi_A(X)$
 supported entirely on 
$A$
\begin{equation}
    \phi(X) ~ \longrightarrow ~~ \Phi_A(X).
\end{equation}
This statement is a precise formulation of subregion duality and has become a cornerstone in our understanding of holography.

A decisive step toward establishing EWR was taken by Jafferis, Lewkowycz, Maldacena, and Suh (JLMS) in  \cite{jafferis2016relative}. They showed that, for states 
$ \rho$, and $\sigma$ corresponded to semiclassical states close to each other, the relative entropy of reduced density matrices on a boundary region 
$A$ equals the relative entropy of the corresponding bulk states restricted to the entanglement wedge 
$\mathcal{E}_A$,
\begin{equation}
    D(\rho_A|\sigma_A) = D(\rho_a|\sigma_a) + O (1/N)
\end{equation}
to leading order in the large 
$N$ expansion. Equivalently, the boundary modular Hamiltonian 
$ K_A = - \log \sigma _A$, when projected onto the code subspace, decomposes as
\begin{equation}
    K_A = \frac{Area(\chi_A)}{4G_N} +K_a^{bulk}+O(1/N).
\end{equation}
This result provided the first precise information-theoretic characterization of subregion duality and identified entanglement, rather than causality, as the fundamental organizing principle behind bulk reconstruction.

The physical significance of the JLMS relation lies in its connection to recoverability in quantum information theory. Relative entropy is monotonic under quantum channels, and saturation of this monotonicity implies the existence of a recovery map. The JLMS equality, therefore, strongly suggests that the encoding of bulk degrees of freedom into the boundary CFT realizes a form of quantum error correction, with the entanglement wedge playing the role of the protected logical region. From this perspective, entanglement wedge reconstruction is not merely a geometric statement, but a structural consequence of how information is redundantly encoded in holography.

A concrete and explicit proof of EWR was subsequently provided by Faulkner and Lewkowycz in \cite{Faulkner:2017}. Building on the JLMS relation, they demonstrated that bulk operators localized within 
$a= \mathcal{E}_A$ can be reconstructed from boundary operators supported on 
$A$ by evolving them under boundary modular flow. This construction confirms that modular flow is the mechanism by which bulk locality arises from boundary entanglement, and it establishes EWR rigorously at leading order in $ 1/N$ expansion.

Despite its conceptual success, modular-flow–based reconstruction has limitations. Its validity relies heavily on the large 
$N$ (semiclassical) approximation where bulk gravity is classical and the JLMS equalities hold exactly at leading order. When quantum gravitational corrections are included, relative entropy equality receives 
$ 1/N$ corrections and boundary modular flow no longer is able to generate a reconstruction for bulk operators. This limitation obscures how EWR should be extended beyond the large 
$N$ regime and motivate the search for reconstruction frameworks that are both explicit and capable of capturing subleading effects.

A related recent perspective on corrections to EWR was developed by Witten \cite{Witten:2026gze}. In the framework of state-dependent geometries introduced by Cao, Cheng, Karthikeyan, Li, and Preskill \cite{cao2026statedependentgeometriesmagicenrichedquantum}, Witten showed that when the area contribution to the generalized entropy is parametrically of order \(1/G_N\), while the bulk entropy is of order one, the violation of EWR can be exponentially small in \(G_N\) compared with the state-dependent variation of the area term. This provides an important distinction between gravitational backreaction and the accuracy of bulk reconstruction: the geometric area function may receive parametrically larger state-dependent corrections than those associated with the failure of reconstruction. The analysis is formulated in terms of an effective area function rather than an area operator. Our approach is complementary: rather than estimating the size of the reconstruction error, we provide an explicit reconstruction channel beyond the leading semiclassical limit, using the twirled Petz map to systematically incorporate subleading corrections to the boundary representation of bulk operators.

A powerful alternative framework emerges from viewing holography explicitly as operator algebra quantum error correction \cite{almheiri2015bulk, dong2016reconstruction}. In this language, EWR is understood as the existence of a dual of a recovery channel that reconstructs bulk operator algebras from boundary subalgebras. In quantum information theory, a canonical such recovery channel is provided by the Petz map, which reconstructs states after a quantum channel whenever relative entropy is preserved. For a channel 
$\mathcal{E}$ and reference state 
$\sigma$, the Petz recovery map is given by
\begin{equation}\label{14}
    \mathcal{R}(.)= \mathcal{P}_{\sigma, \mathcal{E}}(.)= \sigma^{1/2} \mathcal{E}^* \Big(\mathcal{E}(\sigma)^{-1/2} (.)\mathcal{E}(\sigma)^{-1/2}\Big)\sigma^{1/2}
\end{equation}
Recently, it was shown in \cite{Bahiru:2022ukn} that applying the Petz map to the holographic encoding by taking an appropriate basis for the code subspace using the "Reeh-Schlieder" theorem reproduces the same leading-order entanglement wedge reconstruction formula previously obtained using modular flow. This establishes the Petz map as an alternative—and conceptually cleaner—derivation of EWR directly from quantum information–theoretic principles. Crucially, unlike modular flow, the Petz map naturally generalizes to situations where relative entropy equality is only approximate, where one can use the twirled Petz map, which defines an optimal recovery channel even when JLMS receives 
$1/N$ corrections. This makes the twirled Petz map a natural and systematic tool for accessing subleading corrections to entanglement wedge reconstruction, which are inaccessible in modular-flow-based approaches.

In this work, we exploit the twirled Petz map within the framework of operator algebra quantum error correction to derive explicit bulk reconstruction formulas beyond the large 
$N$ limit. Our construction reproduces known semiclassical results while providing controlled access to quantum corrections, thereby extending EWR into the genuinely quantum gravitational regime. This approach clarifies the operational meaning of subregion duality and highlights the central role of quantum information–theoretic recovery maps in the emergence of bulk locality from boundary entanglement.
Beyond the leading approximation, we have obtained an explicit bi-local correction controlled by a subleading Petz kernel
\begin{equation}
    \begin{aligned}
        \Phi_A(X) =& \int_{-\infty}^{\infty} ds~ \int _{x_A \in  A} dx_A~K^{(0)}_{Petz}(X|x_A,s)~O(s,x_A)
        \\& ~~+ \frac{\lambda}{N} \int_{-\infty}^{\infty} ds_1~ \int_{-\infty}^{\infty} ds_2 \int _{x^1_A, x_A^2 \in  A}  dx_A^1 dx_A^2 ~ K^{(1)}_{Petz}(X|x_A^1,x_A^2,s_1,s_2)~ O(s_1,x_A^1)~O(s_2,x_A^2).
    \end{aligned}
\end{equation}
where 
\begin{equation}
    O(s,x_A) = e^{i K_A s } O(x_A) e^{-i K_A s },
\end{equation}
is the modular evolution of the operator algebra on $A$ and $K^{(0)}_{Petz}(X|x_A,s)$, and $K^{(1)}_{Petz}(X|x_A^1,x_A^2,s_1,s_2) $ are the leading order and first subleading order Petz kernels respectively.
The resulting expression offers a concrete implementation of holographic quantum error correction beyond the generalized free field regime, making manifest how bulk interactions are perturbatively encoded in the modular structure of boundary operators. 

The remainder of this paper is organized as follows. In Section 2, we review our setup, which includes the quantum error correction and a brief review of the holographic dictionary and global HKLL reconstruction. In Section 3, we discuss the EWR and provide the relevant formula of the Petz map and twirled Petz map for the holographic setup. 
In Section 4, we discuss the appropriate choice of the basis for the code subspace that simplify the calculation, which is based on the Reeh-Schlieder theorem.
In Section 5, we reproduce the result in \cite{Bahiru:2022ukn}, which is the leading order formula for the EWR, and finally in Section 6, we derive the first subleading correction in terms of the bi-locals modular flowed operator in region $A$.

\section{Setup}

\subsection{Brief review on Quantum Error Correction}

The mathematical framework for quantum error correction (QEC) contains an "isometric embedding" of a small code subspace $\mathcal{H}_{code}$ into a larger Hilbert space  $\mathcal{H}_{phy}$.
\begin{equation}
    V~: ~~  \mathcal{H}_{code}~ \longrightarrow  \mathcal{H}_{phy}
\end{equation}
where $ V^\dagger V = I$. 
In the general theory of QEC, the noise model is described by a quantum channel $\mathcal{E}$, the Kraus operators 
of the channel $\mathcal{E}$ are corresponding to the set of errors.
As one of the convenient descriptions of a quantum channel, instead of the Kraus representation, consider that a natural picture of open quantum dynamics treats the system as coupling to an environment via a joint unitary. After evolution, the environment is no longer accessible, so we trace it out to obtain the system’s state. In fact, any quantum channel can be implemented this way: there is an environment prepared in $ \sigma_{en}$  and a unitary $U$  such that
\begin{equation}\label{1}
     \mathcal{E} (\rho) = \Tr_{en} \big( U( \rho\otimes \sigma _{en} ) U^\dagger\big),
 \end{equation}

Here, the complete error correction procedure is done by another quantum channel $\mathcal{R}$ called "Recovery Channel". The code subspace can be corrected if we require for every state $\rho$ whose support lies in the 
$\mathcal{H}_{code}$
\begin{equation}\label{5}
    \mathcal{R} \circ \mathcal{E} (\rho)=\rho
    \qquad \forall \rho = P_{code}~ \rho ~P_{code},
 \end{equation}
 that $P_{code}$ is the projection onto the code subspace. In other words
 \begin{equation}
     \mathcal{R} \circ \mathcal{E} (\rho_{code})=\rho_{code}
 \end{equation}
while $ \rho_{code} \sim   P_{code}~ \rho ~P_{code}$, and 
it means that, equivalently, the recovery channel projects out the redundant portions of the physical Hilbert space, leaving the matrix code exactly on the remaining support $\mathcal{H}_{phy}$ \cite{Balasubramanian:2023xdp}.

One might be interested to consider the physical system instead of code subspace. In such a case, if we take 
\begin{equation}
    V~: ~\mathcal{H}_{system} ~\longrightarrow~ \mathcal{H}_{phy}
\end{equation}
as the isometry that embeds the $\mathcal{H}_{system}$ into $\mathcal{H}_{phy}$, we can rewrite (\ref{5}) as the following
\begin{equation}
\mathcal{R} \circ \mathcal{E} (V\rho V^\dagger)=V\rho V^\dagger
    \qquad \forall \rho \in S(\mathcal{H}_{system})
\end{equation}
that is equivalent to having
$ \mathcal{E}'$
 and $\mathcal{R}'$ such that 
$\mathcal{R}' \circ \mathcal{E}' (\rho )=\rho $
where \cite{beny2009quantum}
\begin{equation}
    \begin{split}
        \mathcal{E}'(.)& = \mathcal{E}(~V(.)V^\dagger~)
        \\
        \mathcal{R}'(.) &=V^\dagger ~\mathcal{R}(.)~ V.
    \end{split}
\end{equation}

 As mentioned, a general quantum channel 
$ \mathcal{E}: S(\mathcal{H}_A) \rightarrow S(\mathcal{H}_B)$ is reversible with respect to the set
$ \mathcal{Q} \subseteq S(\mathcal{H}_A)$ if there exists another quantum channel $\mathcal{R}$, called the "Recovery channel", such that 
\begin{equation}
    \mathcal{R} \circ \mathcal{E} (\rho)=\rho
    \qquad \forall \rho \in \mathcal{Q}.
\end{equation}
Reversibility of quantum channels has been extensively studied in \cite{jenvcova2006sufficiency, mosonyi2004structure, petz1986sufficient, petz1988sufficiency}.  It is closely linked to the behavior of the quantum relative entropy of states under the action of  $\mathcal{E}$. The relative entropy between two states $\rho$ and $\sigma$  is defined as $ D(\rho | \sigma) = \Tr (\rho \log \rho - \rho \log \sigma ) $ and it is a measure of distinguishability between two quantum states. 
The most important theorem related to this quantity
known as "monotonicity of relative entropy" or the "data processing inequality" \cite{lindblad1975completely, uhlmann1977relative}
\begin{equation}
    D(\rho | \sigma) \geqslant D ( \mathcal{E}(\rho)| \mathcal{E}(\sigma)).
\end{equation}
It has been shown in \cite{petz2003monotonicity,hayden2004structure} that there exists a quantum channel $\mathcal{R}$ such that for all states $\rho \in \mathcal{Q}$, $  \mathcal{R} \circ \mathcal{E} (\rho)=\rho$  if and only if 
$D(\rho | \sigma) = D ( \mathcal{E}(\rho)| \mathcal{E}(\sigma))$
for all $ \rho , \sigma \in \mathcal{Q}$.

Moreover, the explicit form of the quantum channel $\mathcal{R}$ for the set of states 
$\{\mathcal{E}(\rho) | \forall \rho \in \mathcal{Q}\}$ has been given by Petz and his co-workers \cite{hayden2004structure}. It is given as a function of a quantum state 
$\sigma \in S(\mathcal{H}_A)$ and the channel $\mathcal{E}$ itself as
\begin{equation}\label{14}
    \mathcal{R}(.)= \mathcal{P}_{\sigma, \mathcal{E}}(.)= \sigma^{1/2} \mathcal{E}^* \Big(\mathcal{E}(\sigma)^{-1/2} (.)\mathcal{E}(\sigma)^{-1/2}\Big)\sigma^{1/2}
\end{equation}
$\mathcal{E}^*$ is the Hilbert-Schmidt adjoint of $\mathcal{E}$ defined as 
\begin{equation}\label{dual}
    Tr(\rho\mathcal{E}^* (O) )=Tr(\mathcal{E}(\rho) O) \qquad \forall \rho, O.
\end{equation}
$\mathcal{P}_{\sigma, \mathcal{E}}$ is known as "Petz recovery channel". This result has also been independently obtained by Barnum and Knill in \cite{barnum2002reversing}. 

In practice, exact reversibility rarely holds, motivating the study of approximate recovery.  One can ask if there exists an approximate recovery channel that the recovered state is just close to the state $\rho $, 
\begin{equation}
   |\mathcal{R}_ \epsilon \circ \mathcal{E} (\rho) - \rho| < \epsilon  
\end{equation}
where $ \mathcal{R}_ \epsilon  $ is approximate version of recovery channel.
In \cite{wilde2015recoverability, sutter2016strengthened}, it was shown that for any two states $\rho$ and $\sigma$ and channel $\mathcal{E}$, there exists a recovery channel $\mathcal{R}$ such that 
$ \mathcal{R}\circ \mathcal{E} (\sigma)=\sigma $ and
\begin{equation}\label{3}
     D(\rho | \sigma) - D ( \mathcal{E}(\rho)| \mathcal{E}(\sigma))  \geqslant -2 \log F (\rho , \mathcal{R}\circ \mathcal{E} (\rho))
\end{equation}
where 
$ F(\rho , \sigma) :=  \lVert  \sqrt{\rho} \sqrt{\sigma}\rVert _1 $
is the fidelity of $\rho$ and $\sigma$ that measure the closeness of two quantum states. $F(\rho , \sigma) =1  $ if and only if $\rho = \sigma$, then the inequality of \ref{3} will be saturated just in the case of exact reversibility. 
In the following, M. Junge et al \cite{junge2018universal} could show that the $\mathcal{R}$ that satisfies \ref{3} is "universal", that means we can always choose a $\rho$-independent recovery channel.
Furthermore, they could find the explicit expression for the universal recovery map $ \mathcal{R}_{\sigma , \mathcal{E}}$
\begin{equation}\label{4}
    \mathcal{R}_{\sigma , \mathcal{E}}(.)=
    \int _\mathbb{R} dt ~\beta_0 (t) 
    ~\sigma ^{-it/2}~ 
    \mathcal{P}_{\sigma, \mathcal{E}}(\mathcal{E}(\sigma)^{it/2} (.)\mathcal{E}(\sigma)^{-it/2})~
    \sigma ^{it/2}
\end{equation}
where
\begin{equation}
    \beta_0 (t) = \frac{\pi}{2} (( \cosh(\pi t) +1 ))^{-1}
\end{equation}
and $\mathcal{R}_{\sigma , \mathcal{E}}$ is called "twirled Petz map" where $\mathcal{P}_{\sigma , \mathcal{E}}$
is the Petz recovery channel. In the case of exact reversibility, the expression \ref{4} is equal to the Petz map.

The formulation above is called the "standard model for QEC". It involves a triplet 
$ (\mathcal{R}, \mathcal{E}, \mathcal{H}_{code})$ where $\mathcal{H}_{code}$ is a subspace of a Hilbert space
$ \mathcal{H}_{phy}= \mathcal{H}_{code} \oplus \mathcal{H}_{code}^ \perp $ and $\mathcal{R}, \mathcal{E}$ are quantum channels on $ \mathcal{B}(\mathcal{H}_{phy})$. The subspace $ \mathcal{H}_{code}$ is said to be correctable for $\mathcal{E}$ and conserved by $\mathcal{R} \circ \mathcal{E}$.

Conventional quantum error correction is typically presented in the Schrödinger picture. While the Schrödinger and Heisenberg pictures are equivalent in ordinary quantum mechanics, this equivalence can fail in quantum field theory. It has been argued that only the Heisenberg picture is gauge invariant, suggesting that QFT should be formulated in the Heisenberg picture \cite{solomon2007new, solomon1999gauge}.

To cast QEC in the Heisenberg picture, let the system evolve under a channel $\mathcal{E}$. For a state $ \rho $ and observable $ O$, the measurement outcome after the evolution is $ \Tr\big(\mathcal{E}(\rho) O \big)$ . Requiring identical statistics in the Heisenberg picture $\Tr \big(\mathcal{E}^*(O) \rho \big) $, we evolve observables via the dual channel $\mathcal{E}^*$, defined in \eqref{dual}, and the trace preservation of $\mathcal{E}$ is equivalent to the requirement that $\mathcal{E}^*$ is unital, $\mathcal{E}^* (I)=I$.

The conservation of a state by 
$\mathcal{R} \circ \mathcal{E}$ implies that in the Heisenberg picture for all  the operators $O \in \mathcal{L}(\mathcal{H})$ we have
\begin{equation}
    P_{code}~ ( \mathcal{R} \circ \mathcal{E})^* (O)~ P_{code} =  P_{code} ~\mathcal{E}^* \circ \mathcal{R}^*(O)~ P_{code}= P_{code}~ O ~P_{code}.
\end{equation}
Consider that observables can be characterized by a family of operators $\{ X_a\}$. If for every $X_a$ there exists $Y_a$ such that 
\begin{equation}
    X_a=\mathcal{E}^* (Y_a),
\end{equation}
to correct for the errors induced by $\mathcal{E}$, we need the channel $\mathcal{R}$ maps each $X_a$ to one of the operators $Y_a$ through
\begin{equation}
    \mathcal{R}^* (X_a)=Y_a,
\end{equation}
so that 
\begin{equation}
    ( \mathcal{R} \circ \mathcal{E})^* (X_a)= ( \mathcal{E}^* \circ \mathcal{R}^*)(X_a)=X_a.
\end{equation}
In such a case, we will say that $X_a$ is correctable for $\mathcal{E}$ and conserved by $\mathcal{R} \circ \mathcal{E}$.
Therefore, for a given noise model, the conservation of a state by a given noise model implies the conservation of all of its observables. One can think about a natural generalization of the theory of QEC by deducing this strong requirement. Instead of entire observables, we can assume the conservation of just a selecting set of them. 
Thus, for a noise model $\mathcal{E}$, we say that a set $\mathcal{S}$ of operators on $\mathcal{H}$ is correctable on states in the code subspace if there exists a channel $\mathcal{R}$ such that 
\begin{equation}
     P_{code}~ ( \mathcal{R} \circ \mathcal{E})^* (O) ~  P_{code} =  P_{code}~  \mathcal{E}^* \circ \mathcal{R}^*(O) ~  P_{code}=   P_{code}~  O ~ P_{code}, \qquad \forall O \in \mathcal{S}.
\end{equation}
Here, we shall continue to focus on the conservation of sets of operators that have the structure of an algebra where the result is a new theory referred to as "operator algebra quantum error correction" (OAQEC) \cite{kribs2005unified, beny2007generalization}. 
When we have the conservation of just an algebra $\mathcal{A}$ in the code subspace, the information is not encoded into the entire subspace. It is just encoded into some parts of code induced from the algebra structure of $\mathcal{A}$. 

Moreover, we note that one can find the exact expression of the dual of the recovery channel using the equality \eqref{dual}. For the Petz recovery channel, one can find 
\begin{equation}
    \mathcal{R}^*(.)= \mathcal{P}^*_{\sigma, \mathcal{E}}(.)= \mathcal{E}(\sigma)^{-1/2}  \mathcal{E} \Big( \sigma^{1/2} (.)\sigma^{1/2} \Big)\mathcal{E}(\sigma)^{-1/2} 
\end{equation}
and for the twirled Petz recovery channel, one can obtain
\begin{equation}
    \mathcal{R}^*(.)=\int _\mathbb{R} dt ~\beta_0 (t) ~  \mathcal{E}(\sigma)^{-(1+it)/2} ~ \mathcal{E} \Big( \sigma^{(1+it)/2} (.)\sigma^{(1-it)/2} \Big)~\mathcal{E}(\sigma)^{-(1-it)/2} 
\end{equation}


\subsection{Holographic setup}

In this paper, we consider a large N CFT with a holographic dual defined on $ S^{d-1} \times \mathbb{R}$. We take the CFT in the ground state $ \ket{\Omega}$ whose dual spacetime is $ AdS_{d+1}$ which in global coordinate is 
\begin{equation}
    ds^2 = -(1+r^2)dt^2 + \frac{dr^2}{ 1+r^2}+ r^2 d \Omega _{d-1}^2.
\end{equation}
We set both the radius of $ S^{d-1}$ and of $AdS_{d+1}$ to 1.

The correspondence also involves a duality between fields in the bulk and operators in the boundary CFT.
The bulk scalar field $\phi$ is dual to a scalar primary $O$ of the CFT theory with conformal dimension $ \Delta$ related to the mass of the field $\phi$ by 
\begin{equation}
    \Delta = d/2 + \sqrt{ m^2 + d^2/4}
\end{equation}
and the extrapolate dictionary
\begin{equation}\label{extrapolate}
     O(t,\Omega) = \lim _{r \rightarrow \infty} r ^ \Delta ~ \phi (t, r, \Omega). 
\end{equation}
defines $O$ as the representation of $\phi$ at infinity. However, this is a holographic definition and strictly speaking, it is not a boundary condition, since it connects objects living in different spaces. The left-hand side is a CFT operator acting on the CFT Hilbert space, while the right-hand side is the boundary value of a bulk field.
The correct equality is 
\begin{align}
    \lim_{r\rightarrow \infty} r^{n \Delta} \langle \phi (r,t_1,\Omega_1) \phi (r,t_2,\Omega_2)...  \phi (r,t_n,\Omega_n)&\rangle_{\text{pure AdS}}
    \\ =\bra{\Omega} O(t_1,&\Omega_1)O(t_2,\Omega_2)... O(t_n,\Omega_n) \ket{\Omega}.   
\end{align}

Usually, the interpretation above involves taking a large N limit in the CFT, and to leading order at large N, the bulk theory consists of free fields with the Klein-Gordon equation
 \begin{equation}
    (\Box - m^2)\phi=0.
 \end{equation}
Here, we consider the matter field to be scalar for simplicity.
By solving the Klein-Gordon equation for the field $\phi$, we can find the classical solution for the field configuration which are
\begin{equation}
    f_{nlm}( t,r,\Omega)
\end{equation}
labeled by the quantum numbers $n,l$, and $m$, where $ n \in \{0,1,2,...\}$, $l$ is the total angular momentum of the corresponding mode, and $m$ is related to the other angular quantum numbers needed to specify a mode.
In order to quantize the fields $ \phi$ we associate an annihilation operator $ a_{nlm}$ to each mode $f_{nlm} $ such that 
\begin{equation}
     [a_{nlm}, a^\dagger _{n'l'm'}] = \delta _{nn'} \delta _{ll'} \delta _{mm'},
\end{equation}
and the quantized free field on $ AdS_{d+1}$ is given by 
\begin{equation}
    \phi_0 (t,r,\Omega) = \sum _{nlm} f_{nlm} (t, r, \Omega) a_{nlm} + f^* _{nlm} (t, r, \Omega) a^\dagger _{nlm}.
\end{equation}

The most general bulk theory with Einstein gravity and scalar fields dual to a holographic CFT has an action like this
\begin{equation}
    \begin{split}
        S & = \frac{1}{G_N} \int d^{d+1}y \sqrt{-g} R 
        \\
        & + \int d^{d+1}y \sqrt{-g} ( \nabla_\mu \phi \nabla^ \mu \phi + m^2\phi^2)
        \\
        & + \lambda \sqrt{G_N }~ \int  d^{d+1}y \sqrt{-g} \big( \frac{\phi^3}{3!} + \text{all possible cubic couplings}\big)
         \\
        & + \lambda' ~G_N \int  d^{d+1}y \sqrt{-g} \big( \frac{\phi^4}{4!} + \text{all possible quartic couplings}\big) + ...
    \end{split}
\end{equation}
where $ \lambda,\lambda'$ are $ O(1)$ numbers and $G_N$ denotes the Newton's constant.
For the class of theories discussed above, the extrapolate dictionary \eqref{extrapolate} gives us a way to
relate bulk fields near the boundary to  CFT primary operators \cite{Kajuri:2020vxf}.

We will always work in the regime where bulk geometry is semi-classical. For a CFT state to be dual to a semiclassical geometry the gravitational constant $ G_N \ll l^{d-1}$ where
 $ l$ is the AdS radius.
One of the important conditions in holographic CFT is that the CFT must
have a parameter $ N\gg1$ corresponding to the CFT central charge, which controls the factorization of the correlators of the primary
operators which are dual to bulk fields. 
Note that $N$ is related to the expansion parameter in the CFT.

For example, consider the action of $ \mathcal{N}=4$ super Yang-Mills theory with gauge group $N$ that can be written as 
$ N \Tr L$, where $L$ is a gauge-invariant polynomial in the fields of the CFT theory and their derivatives. $L$ has no explicit factor of $N$. In terms of this normalization, the correlation functions of any single-trace operator have a simple large $N$ scaling. The one-point function is of order $N$, a connected two-pointed function of order 1, and in general a connected $k-$point function of order $ N^{2-k}$. \cite{Witten:2021unn}

In the CFT  side, $N$ is the expansion parameter and related to the gravitational constant is related to $N$  as
\begin{equation}
    G_N = \frac{1}{N^2}.
\end{equation}

The discussion above was for pure AdS. In general, in holographic setup, for any semi-classical asymptotically AdS geometry
$g_{\mu \nu}$ we expect that there will be a dual state $ \ket{\Psi_g}$
\begin{equation}
    g_{\mu \nu} ~\longleftrightarrow ~\ket{\Psi_g}
\end{equation}
where
\begin{align}\label{extdic}
    \lim_{r\rightarrow \infty} r^{n \Delta} \langle \phi (r,t_1,\Omega_1) \phi (r,t_2,\Omega_2)...  \phi (r,t_n,\Omega_n)&\rangle_{\text{g}}
    \\ =\bra{\Psi_g} O(t_1,&\Omega_1)O(t_2,\Omega_2)... O(t_n,\Omega_n) \ket{\Psi_g}.   
\end{align}


\subsection{Boundary representation for the bulk fields}

The extrapolate dictionary \eqref{extrapolate} relates bulk fields near the boundary to CFT operators on the boundary, but it doesn’t specify how to represent fields deep in the interior. Bulk reconstruction aims to identify CFT operators $ \Phi_{CFT}(X)$ that reproduce bulk fields $ \phi(X)$ at arbitrary points in the bulk. In other words, we seek for the representations that satisfy
\begin{equation}
    \langle \phi(X_1) \phi(X_2)\rangle_g = \bra{\Psi_g} \Phi_{CFT}(X_1)\Phi_{CFT}(X_2) \ket{\Psi_g}.
\end{equation}

This problem, originally solved through the so-called HKLL reconstruction, was developed by Hamilton, Kabat,
Lifschytz and Lowe in a series of papers \cite{hamilton2006local, hamilton2006holographic, hamilton2007local, hamilton2008local}
In the CFT theory, the aim is to find the operator that represents the bulk field
at finite N, but  we  try to approximate this in a perturbation series in $ 1/N$
as
\begin{equation}
    \Phi_{HKLL}(X) = \Phi^{(0)}_{HKLL}(X) + \frac{1}{N} \Phi^{(1)}_{HKLL}(X) + \frac{1}{N^2} \Phi^{(2)}_{HKLL}(X) +...
\end{equation}
To obtain the zeroth-order term, they reconstruct bulk gravitational operators in the free (non-interacting) limit from boundary operators. In this approximation, bulk operators are represented as smeared single-trace operators in the CFT as
\begin{equation}
      \Phi_{HKLL}^{(0)} (X)= \int_{bdy}  dt dy~ K(X|t,y)~ O(t,y).
\end{equation}
There is freedom is choosing a smearing function which allows us to put the smearing function in a convenient form. In particular, it can be chosen to have support only on boundary points that are spacelike separated from the bulk point $Y= (r, t, \Omega)$, which is the minimal support achievable.
\begin{figure}[h]
    \centering
    \includegraphics[width=0.3\textwidth] {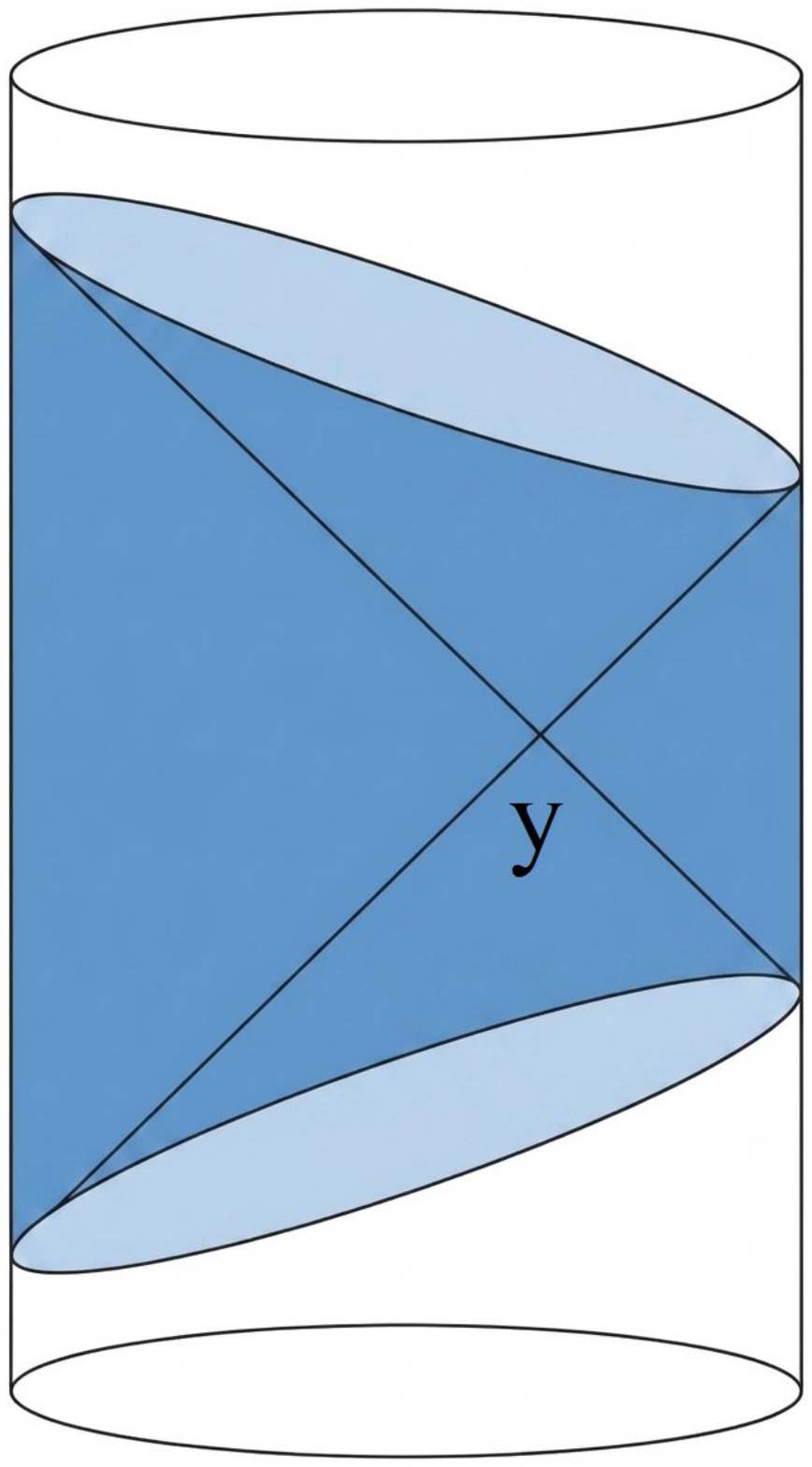}
    \caption{The minimal support of the smearing function for the CFT representation of the bulk field at point $Y$.
    }
    \label{subspace}
\end{figure}

The higher order correction in $ 1/N$ can be done in several ways \cite{kabat2011constructing,kabat2018does, heemskerk2012bulk}.
One is to treat bulk reconstruction as
a boundary value problem, and then one can introduce an appropriate Green’s function and solve the interacting theory order by order. The result is
\begin{equation}\label{HKLL}
    \begin{split}
         \Phi_{HKLL}(X) &= \Phi^{(0)}_{HKLL}(X) + \frac{1}{N} \Phi^{(1)}_{HKLL}(X) +... 
         \\
         & = \int_{bdy}  dt dy~ K(X|t,y)~ O(t,y)
         \\
         & + \frac{\lambda}{N} \int_{bdy} dX' dt_1 dy_1~ dt_2dy_2~ G(X|X') ~K(X'|t_1,y_1)~K(X'|t_2,y_2)~ O(t_1,y_1)O(t_2,y_2)
         \\
         & + ...
    \end{split}
\end{equation}
Note that this is a very non-local operator in the CFT. However, one can choose the support of $G(X|X') $ and $ K(X,y)$ to be non-zero on spacelike separated regions.
As expected, it reproduces the correct bulk two-point function. The smearing function encodes the bulk information into the boundary operator.
In general, the HKLL reconstruction procedure can be extended to other coordinate systems, such as Poincaré coordinates or AdS-Rindler coordinates.


\section{Entanglement Wedge Reconstruction and Quantum Error Correction}

Bulk reconstruction admits many boundary representations of a given low‑energy bulk operator. These boundary operators are distinct on the full CFT Hilbert space, including high‑energy states. The holographic quantum error correction (QEC) proposal interprets this redundancy as equivalence within a low‑energy code subspace. At $ N \rightarrow \infty $ the CFT reduces to a generalized free field (GFF) theory, dual to a free bulk theory, and this redundancy can be viewed as QEC in the sense of \cite{almheiri2015bulk}. However, the code subspace is equivalent to the entire Hilbert space of GFF because the
GFF only contains the low-energy modes, i.e. it is trivial as a quantum error correction
code.

A natural question in AdS/CFT is whether the CFT representation of $ \phi(X) $ can be confined to a subregion of a Cauchy slice $ \Sigma$ that contains $X$. One might expect the required boundary support to shrink as $X$ approaches the boundary. However, \eqref{HKLL} shows that even when $X$ is taken arbitrarily close to the boundary, the reconstruction generically has support over the entire $ \Sigma$ \cite{Bahiru:2022ukn}.
In fact, it is possible to reconstruct bulk operators so that they are supported on smaller
regions on the boundary. The AdS-Rindler reconstruction can be extended to a more general class of bulk regions, the causal wedges of ball-shaped boundary regions. This is called the "causal wedge reconstruction" conjecture. 

 One can go ahead and look at the Rindler Hamiltonian in the AdS-Rindler HKLL reconstruction formulas as the
modular Hamiltonian of the region $A$ that generates the modular flow of operators on $A$.  In \cite{jafferis2016relative}, authors showed
that the boundary modular flow is dual to the bulk modular flow in the entanglement wedge $\mathcal{E}_A$ (which is known as JLMS statement) and conjectured that operators in the entanglement wedge of the region $A$ are the ones
can be constructed on the boundary region $A$ by replacing the Rindler time by the modular
parameter.
The entanglement wedge of a boundary region $A$ is defined as the bulk domain of dependence of any bulk spacelike surface whose boundary is the union of $A$ and the codimension
two extremal area surface of minimal area (more precisely, quantum extremal surface) whose
boundary is $ \partial A$.

To demonstrate the JLMS statement, first, we define the modular Hamiltonian as 
\begin{equation}
    K_\rho := - \log \rho 
\end{equation}
and the relative entropy of two states $\rho$ and $ \sigma$ is 
\begin{equation}
    D(\rho|\sigma) = \Tr (\rho \log \rho) - \Tr (\rho \log \sigma)  = -S(\rho) + \Tr (\rho K_\sigma). 
\end{equation}
The JLMS showed that 
\begin{equation}
    K_A = K_a + \mathcal{A}_{loc}
\end{equation}
where $\mathcal{A}_{loc}$ denotes a bulk operator which is a local integral over the HRT surface $\chi_A$. At leading order in large $N $ limit, we have $\mathcal{A}_{loc} = Area (\chi_A)/4G_N$. Therefore, at leading order in $1/N$ correction, we arrive at 
\begin{equation}
       K_A = K_a + \frac{Area(\chi_A)}{4G_N}.
\end{equation}
In \cite{dong2016reconstruction}, the result is modified and rewritten in the code subspace as 
\begin{equation}\label{JLMSQEC}
    P_{code}~  K_A ~ P_{code} = K_a + \frac{Area(\chi_A)}{4G_N}.
\end{equation}
Moreover, knowing this result, they showed that to order $N^0$ in the large $N$ limit, the relative entropies of the bulk and boundary are equivalent due to the extremality of the HRT surface $\chi_A$.
In other words, consider a Cauchy slice of the boundary CFT and split it into a region $A$ and its complementary region $\bar{A}$. We have 
\begin{equation}
    \mathcal{H}_{CFT} = \mathcal{H}_A \otimes \mathcal{H}_{\bar{A}}.
\end{equation}
and consider the code subspace as 
\begin{equation}
    \mathcal{H}_{code}  \cong \mathcal{H}_{code}^{bulk} = \mathcal{H}_a \otimes \mathcal{H}_{\bar{a}}.
\end{equation}
Consider two bulk states $\rho$ and $\sigma$ that share the same semi-classical geometry. Let us define as the reduced-density matrices on the EW of the boundary region $A$ as
\begin{equation}
    \rho_a = \tr_{\bar{a}} \rho, ~~~~~~~~~~~~~~~ \sigma_a = \tr_{\bar{a}} \sigma.
\end{equation}
The corresponding boundary region reduced density matrices are defined as 
\begin{equation}
    \rho_A = \tr_{\bar{A}} \rho, ~~~~~~~~~~~~~~~ \sigma_A= \tr_{\bar{A}} \sigma.
\end{equation}
Jafferis et al \cite{jafferis2016relative}
demonstrated that 
\begin{equation}
   D(\rho_A|\sigma_A) = D (\rho_a|\sigma_a) + O(1/N).
\end{equation}

Recently, the relation between the bulk and boundary, i.e., the JLMS statement, was used in \cite{dong2016reconstruction} to prove the entanglement wedge reconstruction that interpreted holography as a QEC code. The one-loop result of the JLMS provides the machinery to obtain a leading term in the large N limit of the EWR. It can be done through the Petz recovery channel as the equality between the relative entropies is exact at the leading term.
Moreover, in \cite{faulkner2017bulk}, the authors provide a formula using the modular flow, and recently in \cite{Bahiru:2022ukn}, it has been obtained the modular flow formula has been obtained from the Petz map reconstruction to the leading term at large $N$ limit. 

An important ingredient supporting this is the observation of JLMS that the relative entropy of two states in the boundary region $A$ is equal to the relative entropy of the two
corresponding bulk states in the entanglement wedge up to a subleading correction. Therefore at leading term, by introducing a proper channel that maps the bulk density matrices in $a = \mathcal{E}_A$ to the density matrices in region $A$ as
\begin{equation}
    \rho_A = \mathcal{N} (\rho_a)
\end{equation}
we can use the Petz theorem. At $N \rightarrow \infty$, the JLMS statement guarantees the exact recovery of the information, and for subleading correction, one can use the twirled Petz map \cite{Chen:2019gbt}.

In the language of QEC, in this work, we will be interested in subsystem codes, i.e., quantum codes with complementary recovery, which are the types of codes relevant for entanglement wedge reconstruction in AdS/CFT.


\subsection{Isometric embedding of the code subspace}

In the large $N $ limit of AdS/CFT correspondence, the bulk description is well approximated by semi-classical gravity and the bulk-to-boundary map embeds bulk semi-classical states into dual CFT Hilbert space.
The semiclassical bulk states of interest are built from effective field theory (EFT) excitations with bounded energy density around a chosen gravitational background $g_{\mu\nu}$. These can be organized into a bulk EFT Hilbert space $\mathcal{H}_{bulk,EFT}$. Over the past two decades, many works have examined how this bulk gravity EFT embeds into the boundary CFT \cite{Almheiri:2014lwa,Papadodimas:2012aq, Papadodimas:2013jku, Papadodimas:2015jra}.

We define the bulk code subspace as
\begin{equation}
    \begin{split}
    \mathcal{H}^{bulk}_{code} = &~ \mathcal{A}_{bulk, EFT} \ket{\Omega}
        \\
        =& ~ \text{span} \{ \ket{\Omega}, \phi(X_1)\ket{\Omega}, \phi(X_1)\phi(X_2)\ket{\Omega}, ...\}
    \end{split}
\end{equation}
where $  \mathcal{A}_{bulk, EFT}  = \mathcal{B}( \mathcal{H}_{bulk, EFT} ) $ and the number of bulk field inserted are finite. More generally, one can create the bulk code subspace around any semi-classical geometry by replacing $ \ket{\Omega}$ with the vacuum of that geometry. We will study the bulk reconstruction of the bulk field within this code subspace.
The code subspace is the subset of the full CFT Hilbert space on which a bulk local EFT operator can be represented. States outside this subspace are understood to have lost access to that operator—either because it is absorbed by a large black hole or because the emergent background metric $g$ breaks down.

As in the large $N$ limit, AdS/CFT provides us an isometry of embedding through the global HKLL map $V$: 
\begin{equation}
    V~:~  \mathcal{H}^{bulk}_{code} = \mathcal{H}_{bulk,EFT}~ \longrightarrow ~ \mathcal{H}_{CFT}.
\end{equation}
It has the structure of the QEC code \cite{verlinde2013black, almheiri2015bulk, pastawski2015holographic}. Therefore, there is a subspace 
of the CFT Hilbert space $\mathcal{H}_{code} $ such that 
\begin{equation}
    \mathcal{H}_{bulk,EFT} \cong \mathcal{H}_{code}  \subseteq \mathcal{H}_{CFT}. 
\end{equation}
On the boundary side, we have 
\begin{equation}\label{statecode}
    V \ket{\psi_{bulk}} = \ket{\Psi_{CFT, code}} 
\end{equation}
while $ \ket{\psi_{bulk}}$ and $\ket{\Psi_{CFT, code}}$ are dual to each other.

From the extrapolate dictionary, we have 
\begin{equation}
    \lim_{r\rightarrow \infty} r^\Delta~ V~ \phi(r,t,\Omega)~ V^\dagger  = O^c (t, \Omega)
\end{equation}
where
\begin{equation}\label{pprim}
    O^c (t, \Omega) := P_{code}~ O (t, \Omega) ~P_{code}
\end{equation}
are projected primary fields where $P_{code}$ is projection onto the code subspace. This is a weaker equation with respect to the extrapolate dictionary.
By using this relation, one can reconstruct the global reconstruction of the bulk field restricted to the code subspace as 
\begin{equation}
     \begin{split}
         \Phi_{HKLL}^c(X) &= \Phi^{c,(0)}_{HKLL}(X) + \frac{1}{N} \Phi^{c,(1)}_{HKLL}(X) +... 
         \\
         & = \int_{bdy}  dt dy~ K(X|t,y)~ O^c(t,y)
         \\
         & + \frac{\lambda}{N} \int_{bdy} dX' dt_1 dy_1~ dt_2dy_2~ G(X|X') ~K(X'|t_1,y_1)~K(X'|t_2,y_2)~ O^c(t_1,y_1)O^c(t_2,y_2)
         \\
         & + ...
    \end{split}
\end{equation}

In particular, the ADM Hamiltonian (which is related to the bulk Hamiltonian by using the constraints from the canonical quantization of gravity) is dual to the CFT Hamiltonian. Here, we introduce the code subspace Hamiltonian $\hat{H}^{code}$
as
\begin{equation}\label{ham}
    \hat{H}^{code} = P_{code} H_{CFT} P_{code}
\end{equation}
which is responsible for the time evolution of the projected primaries in the code subspace as 
\begin{equation}\label{timecode}
      O^c (t, \Omega) = e^{i \hat{H}^{code} t} ~ O^c (t=0, \Omega)~  e^{-i \hat{H}^{code} t}.
\end{equation}
More generally, on any static semi-classical geometry, we are able to introduce the code subspace Hamiltonian as in \eqref{ham}. The motivation and discussion are as follows.

As in \cite{Bahiru:2022oas, Bahiru:2023zlc}, starting with the state $ \ket{\Psi_0}$, we define the code subspace as 
\begin{equation}
    \mathcal{H}_{code} = \mathcal{H}_0= \mathcal{A}_{code} \ket{\Psi_0}
\end{equation}
and we name the projection onto this subspace as 
\begin{equation}
    P_{code} = P_0.
\end{equation}
A similar code subspace can be defined for each of the time-shifted states
\begin{equation}
    \begin{split}
        \mathcal{H}_{code} (t)= \mathcal{H}_t & = \mathcal{A}_{code} \ket{\Psi_t}
        \\
        &= \mathcal{A}_{code} e^{-i H_{CFT}t} \ket{\Psi_0}
    \end{split}
\end{equation}
where $ \mathcal{A}_{code} = \mathcal{L}(\mathcal{H}_{code})$
and its corresponding projection is
\begin{equation}
    P_t = e^{-i H_{CFT}t}~ P_0 ~e^{i H_{CFT}t} 
\end{equation}
In general, we have $ P_t \neq P_0$. But, in such a case like $ \ket{\Psi_0} = \ket{\Omega}$ we have $ P_t = P_0$.

Now, let us go back to the projected primaries. From \eqref{pprim}, we have
\begin{equation}
     O^c (t, \Omega) =  P_{code}~ e^{i H_{CFT} t} ~ O (t=0, \Omega)~  e^{-i H_{CFT} t}~ P_{code}
\end{equation}
By using the fact that $ P_{code}=  P_{code}^2$ and the relation 
\begin{equation}\label{comm}
    P_0 ~ e^{i H_{CFT} t} = e^{i H_{CFT} t} ~ P_t
\end{equation}
we reach to 
\begin{equation}
    O^c (t, \Omega) =  P_{code}~ e^{i H_{CFT} t}~ P_t~ O (t=0, \Omega)~ P_t~ e^{-i H_{CFT} t}~ P_{code}
\end{equation}
Now, for the pure AdS when $\ket{\Psi_0} = \ket{\Omega}$
we have
\begin{equation}\label{codeo}
    \begin{split}
         O^c (t, \Omega) =& ~ P_{code}~ e^{i H_{CFT} t}~ P_{code}~ O (t=0, \Omega)~ P_{code}~ e^{-i H_{CFT} t}~ P_{code}
         \\
         =&  ~e^{i \hat{H}^{code} t} ~ O^c (t=0, \Omega)~  e^{-i \hat{H}^{code} t}
    \end{split}
\end{equation}
where again we used the equality $ P_{code}=  P_{code}^2$ and the fact that 
\begin{equation}
     P_{code}~ e^{i H_{CFT} t}~ P_{code} = \sum _{n=0}^\infty \frac{(it)^n}{n!} P_{code}~ H_{CFT}^n~ P_{code} = \exp \Big(it~ P_{code}~H_{CFT}~ P_{code}\Big).
\end{equation}
In the last line, we use the fact that from  \eqref{comm} one can obtain that over the vacuum $P_{code}$ commutes with the $H_{CFT} $.

In the bulk, we have 
\begin{equation}\label{timebulk1}
    \lim _{r\rightarrow \infty} r^ \Delta \phi(r,t,\Omega) = \lim_{r\rightarrow \infty} r^ \Delta ~ e^{i H_{bulk,matter} t} ~ \phi(r,t=0,\Omega)~e^{-iH_{bulk,matter }t}
\end{equation} 
using the constrain of quantum gravity, at the limit where $\lambda \sim O(1)$ and the gravitational coupling is of order $1/N^2$,  we can rewrite it as 
\begin{equation}\label{timebulk}
    \lim _{r\rightarrow \infty} r^ \Delta \phi(r,t,\Omega) = \lim_{r\rightarrow \infty} r^ \Delta ~ e^{i H_{ADM} t} ~ \phi(r,t=0,\Omega)~e^{-iH_{ADM }t}
\end{equation} 
From the weaker version of the extrapolate dictionary, we reach 
\begin{equation}
    O^c(t, \Omega) = e^{i H_{ADM} t} ~ O^c(t=0,\Omega)~e^{-i H_{ADM }t}.
\end{equation}
From \eqref{timebulk1}, one can also get 
\begin{equation}
    O^c(t, \Omega) = e^{i \hat{H} t} ~ O^c(t=0,\Omega)~e^{-i \hat{H}t}.
\end{equation}
where
\begin{equation}
    \hat{H} = H_{bulk, matter}\big( \phi(r , t =0, \Omega) ~\text{on}~ \Sigma(t=0)  \Longrightarrow \Phi_{HKLL}^c(r,t=0, \Omega)\big).
\end{equation}
Note that $ \hat{H}$ acts on the CFT Hilbert space while $ H_{bulk,matter}$ acts on the bulk Hilbert space.
Now, one question that arises is if the code subspace Hamiltonian $ \hat{H}_{code}$ is related to $ \hat{H}$? 
Using the fact that $V$ is an isometry of embedding, by comparing \eqref{timebulk1} and \eqref{timecode} we find that 
\begin{equation}
     \hat{H}^{code} =  \hat{H} + Q
\end{equation}
where $Q$ belongs to the center of the algebra ($ Q\in \mathcal{Z}_{\mathcal{A}_{code}}$) and $ [Q,\hat{H} ] =0$
\footnote{$Q$ can be obtained through the Krylov subspace diagonalization and the method discussed in \cite{Vardian:2026yar}.}.
Therefore 
\footnote{We note the similar behavior of the operator $Q$ that appears here and the canonical conjugation of the time operator in \cite{jensen2025holographic}.}
\begin{equation}
     [\hat{H}^{code} ,\hat{H} ] =0.
\end{equation}

In our discussion since $\mathcal{A}_{code}$ is a factor and its center is trivial, we have
\begin{equation}
    Q = \alpha I
\end{equation}
for some $\alpha \in \mathbb{R}$.


\subsection{Petz map reconstruction restricted to the code subspace}\label{petz}

Consider we have the vacuum state of the CFT, which is dual to the pure AdS spacetime. 
We can always take the smaller subspace as the code subspace as $ \mathcal{H}'_{code} \subseteq \mathcal{H}_{code}$. We can choose it such that $ \ket{\Omega} \in \mathcal{H}'_{code}$ by considering the small amount of backreaction and $ P_{code}'$ is the projection onto the smaller code subspace.
We have 
\begin{equation}
     P_{code}'  P_{code} = P_{code}', ~~~~~~~~~~~~~P_{code} P_{code}' =P_{code}'.
\end{equation}
In order to do the EWR, we take the region $A$ on the Cauchy slice $\Sigma$ at $ t=0$, and we denote its complementary region as $\Bar{A}$. 
Note that the entanglement wedge is $ a=\mathcal{E}_A $. From the EWR, we know that by using the dual of the recovery channel, one can find the CFT representation 
for the operators in $ \mathcal{E}_A$ such that it has support only on region $A$ \cite{cotler2019entanglement}
\begin{equation}
    \phi(X\in a) ~ \longrightarrow ~ \Phi_A(X) = \mathcal{R}^* (\phi(X)).
\end{equation}
If we work in the large $N$ limit, then the recovery channel is the Petz recovery channel for the quantum channel $ \mathcal{N}$ \eqref{14}. By using the  global HKLL isometry of  embedding, one can write an exact form of the quantum channel as 
\begin{equation}\label{N}
    \mathcal{N}(.) = \Tr_{\Bar{A}} \big[ V( . \otimes \sigma_{\Bar{a}})V^\dagger\big] 
\end{equation}
while $\sigma_{\Bar{a}} $ is an arbitrary density matrix of the region $ \Bar{a} = \mathcal{E}_{\Bar{A}}$.
By using this choice of the quantum channel \eqref{N} and $ \sigma_a$ and $ \sigma_{\Bar{a}}$  (in the definition of the Petz recovery channel) to be the maximally mixed states, we will get the "Petz map" as 
\begin{equation}
      \phi(X\in a) ~ \longrightarrow ~ \Phi_A(X) = \Tr _{\Bar{A}} [  P_{code}'] ^{-1/2}~~  \Tr _{\Bar{A}} [  P_{code}'~ \Phi_{HKLL}(X)~ P_{code}']  ~~ \Tr _{\Bar{A}} [  P_{code}'] ^{-1/2}.
\end{equation}
If we want to consider the $1/N$ correction to the EWR, because of the $1/N$ correction to the JLMS statement \cite{jafferis2016relative}, we should consider the twirled Petz recovery channel, and by choosing again $ \sigma_a$ and $ \sigma_{\Bar{a}}$ to be maximally mixed state for the same definition of the quantum channel \eqref{N}, we have "twirled Petz map" as 
\begin{align}
    \phi(X\in a) ~ \longrightarrow & ~
    \\
    \Phi_A(X) = &\int _{\mathbb{R}} ds \beta_0(s) \Tr _{\Bar{A}} [  P_{code}'] ^{-(1+is)/2}~  \Tr _{\Bar{A}} [  P_{code}'~ \Phi_{HKLL}(X)~ P_{code}']  ~ \Tr _{\Bar{A}} [  P_{code}'] ^{-(1-is)/2}.
\end{align}
We note that to leading order in $1/N$ expansion, we just keep the leading term of the global HKLL operators while in $1/N$ correction, we should keep the next-to-leading terms of the global HKLL operators in \eqref{HKLL}.

In order to precisely find $\Phi_A$, we need to deal with the terms as below
\begin{equation}
    \Tr _{\Bar{A}} [  P_{code}'~ \Phi_{HKLL}(X)~ P_{code}'].
\end{equation}
We can rewrite it as 
\begin{equation}
    \Tr _{\Bar{A}} [  P_{code}'~ P_{code}~ \Phi_{HKLL}(X)~ P_{code}~ P_{code}'].
\end{equation}

Now, one relevant question that arises in the calculation is whether 
\begin{equation}
    P_{code}~ \Phi_{HKLL}(X)~ P_{code}  \overset{?} {=}  \Phi^c_{HKLL}(X).
\end{equation}
In general, the answer might be no, the reason is that by considering $1/N$  correction even over the pure AdS, the geometry will have at least a small amount of backreaction, and the geometry will get time-dependent.  Therefore, $ P_t$ is not the same as $ P_0 = P_{code}$ and for the small backreaction we can consider $ P_t \sim P_0 + O(1/N)$.

Let us check term by term in perturbation theory. The first terms match since from \eqref{codeo} we have 
\begin{equation}
    P_{code}~ O(t, \Omega)~ P_{code} = ~e^{i \hat{H}^{code} t} ~ O^c (t=0, \Omega)~  e^{-i \hat{H}^{code} t} 
\end{equation}
The second term is slightly different. In general, we cannot say that
\begin{equation}\label{subl}
    P_{code}~ O(t_1, \Omega_1)~ O(t_2, \Omega_2)~ P_{code} \neq ~ e^{i \hat{H}^{code} t_1} ~ O^c (t=0, \Omega_1)~  e^{-i \hat{H}^{code} t_1} ~ e^{i \hat{H}^{code} t_2} ~ O^c (t=0, \Omega_2)~  e^{-i \hat{H}^{code} t_2}.
\end{equation}
However, we have from the extrapolate dictionary \eqref{extdic}
\begin{equation}
   \begin{aligned}
\lim_{r\to\infty}r^{2\Delta}~ \langle\phi(r,t_1,\Omega_1)~&\phi(r,t_2,\Omega_2)\rangle_g\\
&=\langle\Psi_g|O(t_1,\Omega_1) O(t_2,\Omega_2)|\Psi_g\rangle=\langle\Psi_g|P_{code}O(t_1,\Omega_1) O(t_2,\Omega_2)P_{code}|\Psi_g\rangle.
\end{aligned} 
\end{equation}
since we have $ |\Psi_g\rangle \in \mathcal{H}_{code}$. We can write this as
\begin{align*}
    \langle\Psi_g|P_{code} & O(t_1,\Omega_1)  O(t_2,\Omega_2)P_{code}|\Psi_g\rangle
    \\
    &=\langle\Psi_g|P_{code}O(t_1,\Omega_1)(P_{code}+1-P_{code}) O(t_2,\Omega_2)P_{code}|\Psi_g\rangle\\
    &=\langle\Psi_g|P_{code}O(t_1,\Omega_1) P_{code}O(t_2,\Omega_2)P_{code}|\Psi_g\rangle+\langle\Psi_g|P_{code}O(t_1,\Omega_1) (1-P_{code})O(t_2,\Omega_2)P_{code}|\Psi_g\rangle
\end{align*}
This can be interpreted as follows: when $O$ acts on a state in the code subspace, the result is a general state in the CFT Hilbert space. The first term keeps the state in the code subspace, which is the generalized free field sector for large $N$, so it acts as a free propagation in the code subspace up to $O(\frac{1}{N})$. The second term includes the part that is outside of the code subspace, and therefore looks like an interacting term of  $O(\frac{1}{N})$. Since we want to keep everything up to order $O(\frac{1}{N})$, and the HKLL reconstruction already has a factor of $1/N$ behind these operators, we can ignore the second term and get the desired result
$e^{i \hat{H}^{code} t_1} ~ O^c (t=0, \Omega_1)~  e^{-i \hat{H}^{code} t_1} ~ e^{i \hat{H}^{code} t_2} ~ O^c (t=0, \Omega_2)~  e^{-i \hat{H}^{code} t_2}$ inside the expectation value.
Therefore, as we will explain in the next chapters for the second term, it is enough to keep the code subspace Hamiltonian up to the GFF term ($ \hat{H}^{code} \sim H_{GFF}$ ). By assuming this and considering the fact that on the Cauchy slice $t=0$ of the bulk theory, the interacting fields are equal to free fields, we can recover some kind of equality in \eqref{subl} as 
\begin{equation}
\begin{aligned}
       \frac{1}{N}&  P_{code}~ O(t_1, \Omega_1) ~ O(t_2, \Omega_2)~ P_{code} 
       \\ &= \frac{1}{N}~ e^{i \hat{H}^{GFF} t_1} ~ O^c (t=0, \Omega_1)~  e^{-i \hat{H}^{GFF} t_1} ~ e^{i \hat{H}^{GFF} t_2} ~ O^c (t=0, \Omega_2)~  e^{-i \hat{H}^{GFF} t_2} + O(1/N^2).
\end{aligned}
\end{equation}


\section{Appropriate basis for the code subspace}\label{appbasis}

The Hilbert space of the effective field theory on the bulk side has the structure of the Fock space. 
As a result, the code subspace $ \mathcal{H}_{code} \cong \mathcal{H}_{bulk,EFT}$ has a Fock space structure. 

In this case, since the bulk fields are not free, But
one can use the fact that in quantum field theory, in an interacting theory where the Hamiltonian $H$ differs from the free Hamiltonian $H_0$, the Heisenberg equation of motion is  still satisfied, and we can write the interacting field as 
\begin{equation}
    \phi(x,t) = \int \frac{d^3p}{(2\pi )^3} \frac{1}{\sqrt{2\omega_p}}\Big( a_p(t) e^{-ipx} + h.c.\Big)
\end{equation}
$ a_p(t)$ and $ a_p^\dagger(t)$ are the interacting creation and annihilation operators in the theory of any fixed time $t$ and we have $ [ a_p(t),  a_{p'}^\dagger(t)] = (2\pi)^3 \delta^3(p-p')$. Therefore, the Fock space is the same at every time due to the time-translational invariance. 

Finally, we can define the interacting modes to be equal to free modes at any fixed time $ t=t_0 $ as $  a_p(t = t_0) :=  a_p$ and as a result
\begin{equation}
    \phi(x,t_0) = \phi_0(x, t_0)
\end{equation}
where $\phi_0(x, t) $ is the free field.

By repeating the discussion above  for the scalar matter fields on the  AdS spacetime, we have 
\begin{equation}
    \phi(r,t,\Omega) = \sum_{nlm} f^I_{nlm}(r, t, \Omega) a_{nlm}(t) + h.c.
\end{equation}
and by setting 
\begin{equation}
    a_{nlm}(t_0=0) = a_{nlm}
\end{equation}
we reach $ \phi(r, t_0=0, \Omega) = \phi_0(r, t_0=0, \Omega)$.
From the extrapolate dictionary, we arrive to
\begin{equation}\label{mapmodes}
    V ~ a_{nlm}~ V^\dagger = O_{nlm},~~~~~~~~~V ~ a^\dagger_{nlm}~ V^\dagger  = O^\dagger_{nlm}.
\end{equation}
Therefore, $ [O_{nlm}, O^\dagger_{n'l'm'}] = \delta_{nn'} \delta_{ll'} \delta_{mm'}$, and 
\begin{equation}
    O^c(t=0, \Omega) = \sum_{nlm} g_{nlm}(r, t=0, \Omega) O_{nlm}(t) + h.c.
\end{equation}
where $g_{nlm}(r, t=0, \Omega) = \lim_{r \rightarrow \infty} r^\Delta ~f_{nlm}(r, t=0, \Omega)$. Moreover we have 
\begin{align}
     O^c(t, \Omega) =& \sum_{nlm} g_{nlm}(r, t=0, \Omega) ~  e^{i \hat{H}^{code} t}~O_{nlm} ~ e^{-i \hat{H}^{code} t} + h.c.
     \\
     =&  \sum_{nlm} g_{nlm}(r, t=0, \Omega) ~  e^{i \hat{H} t}~O_{nlm} ~ e^{-i \hat{H} t} + h.c.
\end{align}

One can write the code subspace as
\begin{equation}
    \mathcal{H}_{code} = \text{span} \{ \ket{\{i_{nlm}\}} :=  \Pi _{nlm} (O_{nlm}^\dagger)^{i_{nlm}} \ket{\Omega}\}
\end{equation}
where $ \ket{\Omega}$ is the global vacuum, and it can be defined as $ O_{nlm} \ket{\Omega} =0 , \forall ~nlm $, and $ i_{nlm} \ge 0$. One can also put a cut-off on $i_{nlm}$. 

In order to do the EWR, we need to deal with terms like $ \Tr_{\bar{A}} [P_{code}]$ and  $ \Tr_{\bar{A}} [P_{code} \mathcal{F}(O) P_{code}]$ where $ \mathcal{F}(O)$ is a function of the projected primaries, and thus first, we need to choose a basis for the code subspace.
As it is explained in detail in \cite{Bahiru:2022ukn}, the usual basis of the code subspace, i.e. $ \ket{\{i_{nlm}\}}$ is inconvenient in order to proceed with the calculation.

In \cite{Bahiru:2022ukn}, the authors define the code subspace as the span of states created by acting with the operator algebra of a special subregion on the corresponding semi-classical state. In other words, the "Reeh-Schlieder" theorem in relativistic QFT has been used.

The Reeh-Schlieder theorem states that in a QFT in a Minkowski spacetime $\mathcal{M}$ with a Hilbert space $\mathcal{H}$ and the vacuum $ \ket{\Omega} \in \mathcal{H}$, there is a bounded algebra of local operators $ \mathcal{A}_{\mathcal{U}}$ for a small open set $ \mathcal{U} \subset \mathcal{M}$ such that every arbitrary state in $\mathcal{H}$ can be approximated by $ \mathcal{A}_\mathcal{U}\ket{\Omega}$.
In the Tomita-Takesaki theory language, it means that the vacuum is a cyclic and separating vector for the field algebra corresponding to any open set $\mathcal{U}$ in the Minkowski spacetime.
This is the key feature of the relativistic QFT that introduces a novel basis for the code subspace, i.e., by using this theorem, one can construct the code subspace by acting on the global vacuum with the operator algebra on the boundary region $A$
\begin{equation}
    \mathcal{H}_{code} = \mathcal{L}(  \mathcal{H}_{A})\Big|_{\mathcal{A}_{code} \cap \mathcal{A}_{A}} \ket{\Omega}  
\end{equation}
while 
\begin{equation}
    \mathcal{L}(  \mathcal{H}_{A})\Big|_{\mathcal{A}_{code} \cap \mathcal{A}_{A}} = \{ O^c(t=0, y) | ~\forall~ y \in A \}.
\end{equation}
This approach simplifies the calculation of the (twirled) Petz map reconstruction and makes the process more manageable.

Let us choose a basis for the operator of the region $ A$ and $ \bar{A} $ restricted to the code subspace as 
$ \{A_\nu ^c\}$ and $ \{\bar{A}_\nu ^c\}$ respectively while 
\begin{equation}
    A_{-\nu} ^c = (A_\nu ^c)^\dagger.
\end{equation}
One can take a basis for the code subspace at large $N$ as 
\begin{equation}
    \ket{\{j_\nu, \Delta_\nu\}} = \Pi _{\nu \geq 0} ~ (A_\nu ^c)^{j_\nu} ~ (A_{-\nu} ^c)^{\Delta_\nu + j_\nu}~ \ket{\Omega}
\end{equation}
where $ j \in \mathbb{N}$, and $ \Delta \in \mathbb{Z}$. 

Consider these basis elements, the calculation will simplify since 
\begin{equation}
    \Tr_{\bar{A}}\ket{\{j_\nu, \Delta_\nu\}} \bra{\{ j'\nu, \Delta'_\nu\}} = \Pi _{\nu \geq 0} ~ (A_\nu ^c)^{j_\nu} ~ (A_{-\nu} ^c)^{\Delta_\nu + j_\nu}~ \Tr_{\bar{A}} (\ket{\Omega} \bra{\Omega}) ~ \Pi _{\nu \geq 0} ~ (A_{\nu} ^c)^{\Delta'_\nu +j'_\nu} ~ (A_{-\nu} ^c)^{j'_\nu}
\end{equation}
is an operator in the region $A$, while $ \Tr_{\bar{A}} (\ket{\Omega} \bra{\Omega}) = \exp (- K_A) $.

In order to go ahead with the (twirled) Petz map calculation, we also need to rewrite the global modes $ O_{nlm} $ and $ O_{nlm}^\dagger$ in terms of the operator algebra in the region $A$ and $\bar{A}$ and, the commutation relation between the operator algebras of the regions $A$ as explained in \cite{Bahiru:2022ukn}.

But practically, whether or not we can explicitly compute all the terms depends on the basis we take, and for an appropriate choice of it, we will be able to find an explicit expression for the (twirled) Petz map reconstruction of a bulk field. Thus, the question is what is a more appropriate choice of the basis for the operator algebra of the regions?

Note that since we might not apply the Bogoliubov transformation directly to the boundary theory to find write $O_{nlm} $ and 
$ O_{nlm}^\dagger$
in terms of the operator algebra in the region 
$A$
and $\bar{A}$ restricted to the code subspace because of the lack of the equation of motion. One needs to use the Bogoliubov transformation between the dual modes on the bulk sides. In the case of the EWR, since we have the JLMS argument in hand, the eigenfunction of the $K_A$ and $ K_{\bar{A}} $ as we will explain below, is a clever choice.

\subsection{Modular flow and modular Hamiltonian eigenfunctions}

As we mentioned, the reduced density matrix of a given region $R$ can be written in terms of its modular Hamiltonian $K_R$ as 
\begin{equation}
    \rho_R =e^{-K_R}.
\end{equation}
$K_R$ generates an automorphism for the operator algebra $ \mathcal{A}_R$ associated to the region $R$ as
\begin{equation}
    a \in \mathcal{A}_R ~ \longrightarrow ~ a_s = e^{iK_R s} ~a ~e^{-iK_R s} \in \mathcal{A}_R.
\end{equation}
It is called the modular flow, which was originally introduced in the context of the algebraic QFT. The Fourier transformation of the operator $a_s$ is given by 
\begin{equation}\label{eigen}
    a_\omega = \int_{-\infty}^\infty ds~ e^{is\omega}~ a_s = \int_{-\infty}^\infty ds~ e^{is\omega}~ e^{iK_R s} ~a ~e^{-iK_R s}
\end{equation}
which are the eigenfunctions of the modular Hamiltonian. In other words
\begin{equation}
    [K_R, a_\omega] = \omega~ a_\omega.
\end{equation}
The set of eigenfunctions of $K_R$ forms a basis for the operator algebra on the region $R$.
We choose the set of $ \{a \in \mathcal{A}_R\}$ such that 
\begin{equation}
    a_\omega^\dagger= a_{-\omega}.
\end{equation}

As a result, in our problem, which is the EWR through the (twirled) Petz map, we can use the set of eigenfunctions of the modular Hamiltonians of both regions $A$ and $ \bar{A}$ as the basis for the operator algebra of the regions. Therefore, we have 
\begin{align}
    [K_A,~ A_\omega] & = \omega~ A_\omega
    \\
       [K_{\bar{A}},~ \bar{A}_\omega] & = \omega~ \bar{A}_\omega
\end{align}
where $ \{A_\omega\} $
and $\{\bar{A}_\omega\}$ are the basis for the operator algebra on region $A$ and $ \bar{A}$ respectively.

On the other hand,  at the large $N $ limit where $ \mathcal{A}_{loc} = Area(\chi_A)/4G_N$, from the JLMS statement,
one can identify the eigenfunctions of the bulk modular Hamiltonian and boundary restricted to the code subspace which for the states in the code subspace reduces to the modular Hamiltonian of the boundary \cite{Akers:2018fow, Dong:2026rce}. In the bulk, one can use the Bogoliubov transformation to write the global modes as a linear combination of the eigenfunctions of the bulk modular Hamiltonians
\begin{equation}
    a_{nlm} = \sum_\omega \alpha^a _{nlm;\omega} A_\omega^{a} + \alpha^{\bar{a}} _{nlm;\omega} \bar{A}_\omega^{\bar{a}}
\end{equation}
and therefore from the extrapolate dictionary and identification of the eigenbasis of the bulk and boundary modular Hamiltonians
\begin{equation}
    A_\omega^{a} \cong A^A_\omega ,~~~~~~~~~~~\bar{A}_\omega^{\bar{a}} \cong \bar{A}^{\bar{A}}_\omega
\end{equation}
we have 
\begin{equation}
    O_{nlm} = \sum_\omega \alpha^a _{nlm;\omega} A^A_\omega + \alpha^{\bar{a}} _{nlm;\omega} \bar{A}^{\bar{A}}_\omega
\end{equation}
in another words, both $a_{nlm}$ and $O_{nlm}$ have the same Bogoliubov coefficients.


\subsection{Explicit expression for the projection onto the code subspace}

In the rest of the paper, we consider the set of eigenbases of the modular Hamiltonian of regions $A$ and $\bar{A}$
as the basis for the operator algebra of the corresponding regions. Therefore, our choice for the basis of the code subspace is 
\begin{equation}
    \ket{\{j_\nu, \Delta_\nu\}} = \Pi _{\nu \geq 0} ~ (A_\nu )^{j_\nu} ~ (A_{-\nu} )^{\Delta_\nu + j_\nu}~ \ket{\Omega}.
\end{equation}
where, we define
\begin{equation}
   A^A_\omega=  A_\omega,~~~~~~~~~~~~~~  \bar{A}^{\bar{A}}_\omega =  \bar{A}_\omega
\end{equation}
Since at $t=0$, the interacting fields are equal to the free fields and $\ket{\Omega}$ is the vacuum of the CFT dual to the pure AdS, and can be decomposed in terms of different modes we have
\begin{equation}
     \ket{\{j_\nu, \Delta_\nu\}} = \otimes_{\nu \geq 0} \ket{j_\nu, \Delta_\nu}= \otimes _{\nu \geq 0} ~ (A_\nu )^{j_\nu} ~ (A_{-\nu} )^{\Delta_\nu + j_\nu}~ \ket{\Omega}.
\end{equation}
One should be careful that, although this set of vectors spans the entire code subspace, they are not orthogonal since we have 
\begin{equation}\label{inner}
    \langle j, \Delta\ket{ j', \Delta'}  \sim \delta _{\Delta,\Delta'}  \sum _{n= max\{0, -\Delta\}} e^{-2\omega n} \sqrt{\frac{(n+j+\Delta)!}{(n+\Delta)!}}  \sqrt{\frac{(n+j'+\Delta')!}{(n+\Delta')!}} 
\end{equation}
which is proportional to $ \delta_{\Delta,\Delta'}$ instead of $ \delta _{j,j'}\delta _{\Delta,\Delta'}$. Here, for simplicity, we just focus on one mode of the Fock space.

To write the projection in this set of vectors, consider a generic vector space $V = \text{span} \{ \ket{v_i}\}$. The "metric tensor" on this subspace is defined as 
\begin{equation}
    G=[g_{ij} = \langle v_i \ket{v_j}]
\end{equation}
and its inverse $G^{-1}= [g^{ij}]$ is called the "inverse metric". As a result, we have the relation
\begin{equation}\label{82}
        \sum _j g^{ij} g_{jk} = \delta ^i _k,~~~~~~~~~~~~~~~~
        \sum_j g_{ij} g^{jk} = \delta ^k _i.
\end{equation}
The projection on the subspace $V_I = \text{span} \{ \ket{v_i}, i \in I \}$ is given by 
\begin{equation}
    P_I = \sum_{i,j \in I} g^{ij} \ket{v_i}\bra{v_j}.
\end{equation}

On the basis that we chose, the projection onto the code subspace is 
\begin{equation}
    P_{code} = \sum _{j,j'} \sum _{\Delta, \Delta'} g^{(j,\Delta);(j',\Delta')} \ket{j, \Delta} \bra{j', \Delta'}. 
\end{equation}
From \eqref{inner}, one can see that the metric tensor $G[g_{(j,\Delta);(j',\Delta')}]$ is block-diagonal and each block is labeled by the parameter $\Delta$. Therefore, we can write
\begin{equation}
    G =\oplus _\Delta G_\Delta = \oplus _\Delta [g_{j,j';\Delta} = \langle j, \Delta\ket{ j', \Delta} ].
\end{equation}
As a result, the inverse metric is block-diagonal as well
\begin{equation}
    G^{-1} =\oplus _\Delta G^{-1}_\Delta = \oplus _\Delta [A_{j,j'}^{\Delta}]
\end{equation}
where $ g^{(j,\Delta);(j',\Delta')} = A_{j,j'}^{\Delta} \delta_{\Delta, \Delta'}$ and 
\begin{equation}\label{84}
\begin{split}
  \sum _{j'}  A_{j,j'}^{\Delta} \langle j', \Delta\ket{ j'', \Delta}  = \delta _{j,j''}
  \\
  \sum _{j'}  \langle j, \Delta\ket{ j', \Delta}  A_{j',j''}^{\Delta}  = \delta _{j,j''}.
\end{split}
\end{equation}
Finally, the projection onto the code subspace can be represented as 
\begin{equation}
      P_{code} = \sum _{\Delta} \sum _{j,j'} A_{j,j'}^{\Delta} \ket{j, \Delta} \bra{j', \Delta}. 
\end{equation}

\subsection{Some relevant calculations}\label{calculation}

In order to do EWR through the recovery channel, by considering the form of the projection onto the code subspace as above, one can proceed with the calculation. During the calculation, one should deal with terms like 
\begin{equation}
    \mathcal{F}(\mathcal{O}) = \Tr_{\bar{A}} [P'_{code}~ \mathcal{O}~P'_{code} ].
\end{equation}
We can  consider that 
$ \mathcal{H}_{code}'=\mathcal{H}_{code}$ or consider the code subspace as the
\begin{equation}
  \mathcal{H}_{code}'= \text{span}\{   \ket{\{j_\nu, \Delta_\nu\}} = \Pi _{\nu \geq 0} ~ (A_\nu)^{j_\nu} ~ (A_{-\nu})^{\Delta_\nu + j_\nu}~ \ket{\Omega}| j \in I_1 \subset \mathbb{N}, \Delta \in I_2 \subset \mathbb{Z} \}
\end{equation}


In general, we have 
\begin{equation}
    \mathcal{F}(\mathcal{O}) = \sum _{\Delta, \Delta'} \sum _{j,j'} \sum _{k,k'} 
   ~A_{j,k}^{\Delta} ~A_{k',j'}^{\Delta'} ~~\bra{k, \Delta} \mathcal{O}\ket{k', \Delta'}~~
   \tr_{\Bar{A}}\ket{j, \Delta} \bra{j', \Delta'}
\end{equation}

\begin{itemize}
    \item \textcolor{red}{ $\mathcal{O} = I$}
    
    \begin{equation}
        \mathcal{F}(I) := \tau_A =  \sum _{\Delta} \sum _{j,j'} A_{j,j'}^{\Delta} (A_\nu^c)^{j} (A_{\nu}^{c,\dagger})^{j+ \Delta} e^{-K^c_A} (A^c_\nu)^{j'+\Delta'} (A_\nu^{c,\dagger})^{j'}
    \end{equation}

    \item  \textcolor{red}{ $\mathcal{O} = A^c_\omega, \text{and} ~ A^c_{-\omega} ~~\omega>0$} 

    Since we have 
    \begin{align}\label{111}
           \bra{ \{k_\nu, \Delta_\nu\}} A^c_\omega \ket{\{k'_\nu, \Delta'_\nu\}}= &\bra{...,\{k_\omega, \Delta_\omega\},...} ...\{(k'+1)_\omega, (\Delta'-1)_\omega\}... \rangle
           \\
           \bra{ \{k_\nu, \Delta_\nu\}} A^c_{-\omega} \ket{\{k'_\nu, \Delta'_\nu\}}= &\bra{...,\{(k+1)_\omega, (\Delta-1)_\omega\},...} ...\{k'_\omega, \Delta'_\omega\}... \rangle,
    \end{align}
    one can arrive at
    \begin{align}
        \mathcal{F}(A^c_\omega)&= A^c_\omega ~ \tau _A
        \\
        \mathcal{F}(A^c_{-\omega}) &= \tau _A ~ A^c_{-\omega}.
    \end{align}

    \item  \textcolor{red}{ $\mathcal{O} = A^c_{\omega_1} A^c_{\omega_2} ~~\omega_{1,2}>0$}

    Since we have 
\begin{align*}
\langle \{k_\nu, \Delta_\nu \} | & A^c_{\omega_1} A^c_{\omega2} | \{k'_\nu, \Delta'_\nu \} \rangle
\\
& = \langle \{k_\nu, \Delta_\nu\}|  A^c_{\omega_1}| \{\dots,  k'_{\omega_2}+1, \Delta'_{\omega_2}-1,  \dots\}\rangle 
\\
&= \begin{cases}
\langle \{k_\nu, \Delta_\nu\}|  \{\dots, k'_{\omega_1}+2, \Delta'_{\omega_1}-2, \dots\}\rangle & \omega_1 = \omega_2
\\
\langle \{k_\nu, \Delta_\nu\}| \{\dots, k'_{\omega_1}+1, \Delta'_{\omega_1}-1, \dots, k'_{\omega_2}+1, \Delta'_{\omega_2}-1, \dots\}\rangle & \omega_1\neq \omega_2
\end{cases}
\end{align*}
and therefore, we have 
\begin{equation}
    \mathcal{F}(A^c_{\omega_1} A^c_{\omega_2}) =  A^c_{\omega_1}A^c_{\omega_2}~ \tau _A  .
\end{equation}

\item  \textcolor{red}{ $\mathcal{O} = A^c_{-\omega_1} A^c_{\omega_2} ~~\omega_{1,2}>0$}

Since we have
\begin{align*}
    \langle \{k_\nu, \Delta_\nu \} | & A^c_{-\omega_1} A^c_{\omega_2} | \{k'_\nu, \Delta'_\nu \} \rangle
\\
& = \langle \{\dots,  k_{\omega_1}+1, \Delta_{\omega_1}-1,  \dots\} |  \{\dots,  k'_{\omega_2}+1, \Delta'_{\omega_2}-1,  \dots\}\rangle.
\end{align*}
And as a result, we can find that 
\begin{equation}
      \mathcal{F}(A^c_{-\omega_1} A^c_{\omega_2}) =  A^c_{-\omega_1}A^c_{\omega_2}~\tau _A  .
\end{equation}

     \item  \textcolor{red}{ $\mathcal{O} = A^c_{\omega_1} A^c_{-\omega_2} ~~\omega_{1,2}>0$}

     In order to calculate it, we note that 
     \begin{equation}
         A_{-\nu} \ket{k_\nu, \Delta_\nu} =    A_{-\nu}    A_{\nu}^{k_\nu}  A_{-\nu}^{k_\nu+ \Delta_\nu} \ket{\Omega^c} =  A_{\nu}^{k_\nu}  A_{-\nu}^{k_\nu+ \Delta_\nu+1} \ket{\Omega^c}  + [A_{-\nu}, A_\nu ^{k_\nu}] A_{-\nu}^{k_\nu+ \Delta_\nu} \ket{\Omega^c}
     \end{equation}
where $ [A_{-\nu}, A_\nu ^{k_\nu}] = -k_\nu ~A_\nu ^{k_\nu-1}$. Therefore, we have 
\begin{align*}
 \langle & \{k_\nu, \Delta_\nu \} |  A^c_{\omega_1} A^c_{-\omega_2} | \{k'_\nu, \Delta'_\nu \} \rangle
\\
& = \langle \{k_\nu, \Delta_\nu\}|  A^c_{\omega_1}\Big(| \{\dots,  k'_{\omega_2}, \Delta'_{\omega_2}+1,  \dots\}\rangle  - k_{\omega_2} | \{\dots,  k'_{\omega_2}-1, \Delta'_{\omega_2}+1,  \dots\}\rangle  \Big)
\\
&= \begin{cases}
 \langle \{k_\nu, \Delta_\nu\}|\Big(| \{\dots,  k'_{\omega_2}+1, \Delta'_{\omega_2},  \dots\}\rangle  - k_{\omega_2} | \{\dots,  k'_{\omega_2}, \Delta'_{\omega_2},  \dots\}\rangle  \Big) & \omega_1 = \omega_2
\\
\langle \{k_\nu, \Delta_\nu\}| \langle \{k_\nu, \Delta_\nu\}| \Big(| \{\dots,K'_{\omega_1}+1, \Delta_{\omega_1}-1,...,  k'_{\omega_2}, \Delta'_{\omega_2}+1,  \dots\}\rangle 
\\- k_{\omega_2} | \{\dots,k'-{\omega_1}+1,\Delta_{\omega_1}+1,...  k'_{\omega_2}-1, \Delta'_{\omega_2}+1,  \dots\}\rangle  \Big) & \omega_1\neq \omega_2
\end{cases}  
\end{align*}
which arrive at
\begin{equation}
      \mathcal{F}(A^c_{\omega_1} A^c_{-\omega_2}) = \tau _A ~ A^c_{\omega_1}A^c_{-\omega_2}
\end{equation}

\item  \textcolor{red}{ $\mathcal{O} = A^c_{-\omega_1} A^c_{-\omega_2} ~~\omega_{1,2}>0$}

Since we have 
\begin{align*}
\langle \{k_\nu, \Delta_\nu \} | & A^c_{-\omega_1} A^c_{-\omega_2} | \{k'_\nu, \Delta'_\nu \} \rangle
\\
& = \langle \{\dots,  k_{\omega_1}+1, \Delta_{\omega_1}-1,  \dots\} |A^c_{-\omega_2} | \{k'_\nu, \Delta'_\nu\}\rangle 
\\
&= \begin{cases}
\langle \{\dots, k_{\omega_1}+2, \Delta_{\omega_1}-2, \dots\} | \{k'_\nu, \Delta'_\nu\}\rangle & \omega_1 = \omega_2
\\
 \langle \{\dots, k_{\omega_1}+1, \Delta_{\omega_1}-1, \dots, k_{\omega_2}+1, \Delta_{\omega_2}-1, \dots\} | \{k'_\nu, \Delta'_\nu\}\rangle & \omega_1\neq \omega_2
\end{cases}
\end{align*}
and therefore, we have 
\begin{equation}
    \mathcal{F}(A^c_{-\omega_1} A^c_{-\omega_2}) = \tau _A ~ A^c_{-\omega_1}A^c_{-\omega_2} .
\end{equation}
\end{itemize}

But there are also terms with operators on $\bar{A}$ present in the calculation. To find the trace for these operators, we can use the following relation
\begin{equation}
    \bar{A}_\nu^{c} \ket{\Omega^c}=e^{-K^c_A/2}~ A_{-\nu}^{c}~e^{K^c_A/2}\ket{\Omega^c}.
\end{equation}
By considering $[K^c_A,A_{\nu}^{c}]=\nu A_{\nu}^{c}$, we have
\begin{equation*}
     \bar{A}_\nu^{c} \ket{\Omega^c}=e^{\nu/2}~ A_{-\nu}^{c}\ket{\Omega^c}.
\end{equation*}
Using this result, we will have

\begin{itemize}
    \item \textcolor{red}{$\mathcal{O}=A_{\omega_1}^{c}\bar{A}_{\omega_2}^{c}~~\omega_{1,2}>0$}
    
    Since we have
    \begin{equation}
        \begin{aligned}
            \langle \{k_\nu, \Delta_\nu \} |  A_{\omega_1}^{c}\bar{A}_{\omega_2}^{c} | \{k'_\nu, \Delta'_\nu \} \rangle & =\langle \{k_\nu, \Delta_\nu \} |  A_{\omega_1}^{c}\bar{A}_{\omega_2}^{c} \otimes _{\nu \geq 0} ~ (A_\nu ^c)^{k'_\nu} ~ (A_{-\nu} ^c)^{\Delta'_\nu + j'_\nu}~| \Omega^c \rangle
\\
& =\langle \{k_\nu, \Delta_\nu \} |  A_{\omega_1}^{c} \otimes _{\nu \geq 0} ~ (A_\nu ^c)^{k'_\nu} ~ (A_{-\nu} ^c)^{\Delta'_\nu + j'_\nu}~\bar{A}_{\omega_2}^{c}| \Omega^c \rangle
\\
& =e^{\omega_2/2}\langle \{k_\nu, \Delta_\nu \} |  A_{\omega_1}^{c} \otimes _{\nu \geq 0} ~ (A_\nu ^c)^{k'_\nu} ~ (A_{-\nu} ^c)^{\Delta'_\nu + j'_\nu}~A_{-\omega_2}^{c}| \Omega^c \rangle
\\
& =e^{\omega_2/2}\langle \{k_\nu, \Delta_\nu \} |  A_{\omega_1}^{c}A_{-\omega_2}^{c} \otimes _{\nu \geq 0} ~ (A_\nu ^c)^{k'_\nu} ~ (A_{-\nu} ^c)^{\Delta'_\nu + j'_\nu}~| \Omega^c \rangle
\\
&~  -e^{\omega_2/2}\langle \{k_\nu, \Delta_\nu \} |  A_{\omega_1}^{c} \otimes _{\nu \geq 0} [A_{-\omega_2}^{c},~ (A_\nu ^c)^{k'_\nu}] ~ (A_{-\nu} ^c)^{\Delta'_\nu + j'_\nu}~| \Omega^c \rangle
        \end{aligned}
    \end{equation}
now we note that
\begin{equation}
    [A_{-\omega_2}^{c},~ (A_\nu ^c)^{k'_\nu}]=-k'_\nu(A_\nu ^c)^{k'_\nu-1}\delta_{-\omega_2,\nu}
\end{equation}
which makes the second term zero due to the tensor product on all values of $\nu$. Therefore
\begin{equation}
    \langle \{k_\nu, \Delta_\nu \} |  A_{\omega_1}^{c}\bar{A}_{\omega_2}^{c} | \{k'_\nu, \Delta'_\nu \} \rangle=e^{\omega_2/2}\langle \{k_\nu, \Delta_\nu \} |  A_{\omega_1}^{c}A_{-\omega_2}^{c} | \{k'_\nu, \Delta'_\nu \} \rangle
\end{equation}
This gives
\begin{equation}
    \mathcal{F}(A_{\omega_1}^{c}\bar{A}_{\omega_2}^{c}) = e^{\omega_2/2}\tau _A ~ A^c_{\omega_1}A^c_{-\omega_2} .
\end{equation}
\item \textcolor{red}{$\mathcal{O}=A_{-\omega_1}^{c}\bar{A}_{\omega_2}^{c}~~\omega_{1,2}>0$}
\begin{align*}
    \langle \{k_\nu, \Delta_\nu \} |  A_{-\omega_1}^{c}\bar{A}_{\omega_2}^{c} | \{k'_\nu, \Delta'_\nu \} \rangle=e^{\omega_2/2}\langle \{k_\nu, \Delta_\nu \} |  A_{-\omega_1}^{c}A_{-\omega_2}^{c} | \{k'_\nu, \Delta'_\nu \} \rangle
\end{align*}
Therefore
\begin{equation}
    \mathcal{F}(A_{-\omega_1}^{c}\bar{A}_{\omega_2}^{c}) = e^{\omega_2/2}\tau _A ~ A^c_{-\omega_1}A^c_{-\omega_2} .
\end{equation}

\item \textcolor{red}{$\mathcal{O}=A_{\omega_1}^{c}\bar{A}_{-\omega_2}^{c}~~\omega_{1,2}>0$}
 \begin{align*}
\langle \{k_\nu, \Delta_\nu \} |  A_{\omega_1}^{c}\bar{A}_{-\omega_2}^{c} | \{k'_\nu, \Delta'_\nu \} \rangle & =\langle \{k_\nu, \Delta_\nu \} |  A_{\omega_1}^{c}\bar{A}_{-\omega_2}^{c} \otimes _{\nu \geq 0} ~ (A_\nu ^c)^{k'_\nu} ~ (A_{-\nu} ^c)^{\Delta'_\nu + k'_\nu}~| \Omega^c \rangle
\\
& =\langle \{k_\nu, \Delta_\nu \} |  A_{\omega_1}^{c} \otimes _{\nu \geq 0} ~ (A_\nu ^c)^{k'_\nu} ~ (A_{-\nu} ^c)^{\Delta'_\nu + k'_\nu}~\bar{A}_{-\omega_2}^{c}| \Omega^c \rangle
\\
& =e^{-\omega_2/2}\langle \{k_\nu, \Delta_\nu \} |  A_{\omega_1}^{c} \otimes _{\nu \geq 0} ~ (A_\nu ^c)^{k'_\nu} ~ (A_{-\nu} ^c)^{\Delta'_\nu + k'_\nu}~A_{\omega_2}^{c}| \Omega^c \rangle
\\
& =e^{-\omega_2/2}\langle \{k_\nu, \Delta_\nu \} |  A_{\omega_1}^{c}A_{\omega_2}^{c} \otimes _{\nu \geq 0} ~ (A_\nu ^c)^{k'_\nu} ~ (A_{-\nu} ^c)^{\Delta'_\nu + k'_\nu}~| \Omega^c \rangle
\\
&~  -e^{-\omega_2/2}\langle \{k_\nu, \Delta_\nu \} |  A_{\omega_1}^{c} \otimes _{\nu \geq 0} ~ (A_\nu ^c)^{k'_\nu} ~[A_{\omega_2}^{c},(A_{-\nu} ^c)^{\Delta'_\nu + k'_\nu}~] ~| \Omega^c \rangle\\
&=e^{-\omega_2/2}\langle \{k_\nu, \Delta_\nu \} |  A_{\omega_1}^{c}A_{\omega_2}^{c} | \{k'_\nu, \Delta'_\nu \} \rangle
\end{align*}
which gives us
\begin{equation}
      \mathcal{F}(A_{\omega_1}^{c}\bar{A}_{-\omega_2}^{c}) = e^{-\omega_2/2} A^c_{\omega_1}A^c_{\omega_2}~\tau _A  .
\end{equation}

\item \textcolor{red}{$\mathcal{O}=A_{-\omega_1}^{c}\bar{A}_{-\omega_2}^{c}~~\omega_{1,2}>0$}
\begin{align*}
    \langle \{k_\nu, \Delta_\nu \} |  A_{-\omega_1}^{c}\bar{A}_{-\omega_2}^{c} | \{k'_\nu, \Delta'_\nu \} \rangle=e^{-\omega_2/2}\langle \{k_\nu, \Delta_\nu \} |  A_{-\omega_1}^{c}A_{\omega_2}^{c} | \{k'_\nu, \Delta'_\nu \} \rangle
\end{align*}
Therefore
\begin{equation}
      \mathcal{F}(A_{-\omega_1}^{c}\bar{A}_{-\omega_2}^{c}) = e^{-\omega_2/2} A^c_{-\omega_1}A^c_{\omega_2}~\tau _A  .
\end{equation}

\item \textcolor{red}{$\mathcal{O}=\bar{A}_{\omega_1}^{c}\bar{A}_{\omega_2}^{c}~~\omega_{1,2}>0$}
\begin{align*}
\langle \{k_\nu, \Delta_\nu \} |  \bar{A}_{\omega_1}^{c}\bar{A}_{\omega_2}^{c} | \{k'_\nu, \Delta'_\nu \} \rangle & =\langle \Omega^c | \otimes _{\nu \geq 0} ~(A_{\nu} ^c)^{\Delta_\nu + j_\nu}~ (A_{-\nu} ^c)^{k_\nu} ~  \bar{A}_{\omega_1}^{c}\bar{A}_{\omega_2}^{c} | \{k'_\nu, \Delta'_\nu \} \rangle
\\
 & =\langle \Omega^c | \bar{A}_{\omega_1}^{c}\otimes _{\nu \geq 0} ~(A_{\nu} ^c)^{\Delta_\nu + j_\nu}~ (A_{-\nu} ^c)^{k_\nu} ~  \bar{A}_{\omega_1}^{c}\bar{A}_{\omega_2}^{c} | \{k'_\nu, \Delta'_\nu \} \rangle
 \\
 &
=e^{\omega_1/2}\langle \Omega^c |A_{-\omega_1}^{c} \otimes _{\nu \geq 0} ~(A_{\nu} ^c)^{\Delta_\nu + j_\nu}~ (A_{-\nu} ^c)^{k_\nu} ~  \bar{A}_{\omega_2}^{c} | \{k'_\nu, \Delta'_\nu \} \rangle
\\
&
=e^{\omega_1/2}\langle \Omega^c | \otimes _{\nu \geq 0} ~(A_{\nu} ^c)^{\Delta_\nu + j_\nu}~ (A_{-\nu} ^c)^{k_\nu} ~  A_{-\omega_1}^{c}\bar{A}_{\omega_2}^{c} | \{k'_\nu, \Delta'_\nu \} \rangle
\\
&
~~+e^{\omega_1/2}\langle \Omega^c | \otimes _{\nu \geq 0} [A_{-\omega_1}^{c},~(A_{\nu} ^c)^{\Delta_\nu + j_\nu}]~ (A_{-\nu} ^c)^{k_\nu} ~  \bar{A}_{\omega_2}^{c} | \{k'_\nu, \Delta'_\nu \} \rangle\\
&
=e^{\omega_1/2}\langle \{k_\nu, \Delta_\nu \} |  A_{-\omega_1}^{c}\bar{A}_{\omega_2}^{c} | \{k'_\nu, \Delta'_\nu \} \rangle
\\& =e^{\omega_1/2}\langle \{k_\nu, \Delta_\nu \} |  A_{-\omega_1}^{c}\bar{A}_{\omega_2}^{c} \otimes _{\nu \geq 0} ~ (A_\nu ^c)^{k'_\nu} ~ (A_{-\nu} ^c)^{\Delta'_\nu + j'_\nu}~| \Omega^c \rangle
\\
& =e^{\omega_1/2}\langle \{k_\nu, \Delta_\nu \} |  A_{-\omega_1}^{c} \otimes _{\nu \geq 0} ~ (A_\nu ^c)^{k'_\nu} ~ (A_{-\nu} ^c)^{\Delta'_\nu + j'_\nu}~\bar{A}_{\omega_2}^{c}| \Omega^c \rangle
\\
&=e^{(\omega_1+\omega_2)/2}\langle \{k_\nu, \Delta_\nu \} |  A_{-\omega_1}^{c} \otimes _{\nu \geq 0} ~ (A_\nu ^c)^{k'_\nu} ~ (A_{-\nu} ^c)^{\Delta'_\nu + j'_\nu}~A_{-\omega_2}^{c}| \Omega^c \rangle
\\
&=e^{(\omega_1+\omega_2)/2}\langle \{k_\nu, \Delta_\nu \} |  A_{-\omega_1}^{c}A_{-\omega_2}^{c} \otimes _{\nu \geq 0} ~ (A_\nu ^c)^{k'_\nu} ~ (A_{-\nu} ^c)^{\Delta'_\nu + j'_\nu}~| \Omega^c \rangle
\\
& -e^{(\omega_1+\omega_2)/2}\langle \{k_\nu, \Delta_\nu \} |  A_{-\omega_1}^{c} \otimes _{\nu \geq 0} [A_{-\omega_2}^{c},~ (A_\nu ^c)^{k'_\nu}] ~ (A_{-\nu} ^c)^{\Delta'_\nu + j'_\nu}~| \Omega^c \rangle
\\
& =e^{(\omega_1+\omega_2)/2}\langle \{k_\nu, \Delta_\nu \} |  A_{-\omega_1}^{c}A_{-\omega_2}^{c} \otimes _{\nu \geq 0} ~ (A_\nu ^c)^{k'_\nu} ~ (A_{-\nu} ^c)^{\Delta'_\nu + j'_\nu}~| \Omega^c \rangle
\\
& =e^{(\omega_1+\omega_2)/2}\langle \{k_\nu, \Delta_\nu \} |  A_{-\omega_1}^{c}A_{-\omega_2}^{c} | \{k'_\nu, \Delta'_\nu \} \rangle
\end{align*}
and we get
\begin{equation}
    \mathcal{F}(\bar{A}^c_{\omega_1} \bar{A}^c_{\omega_2}) = e^{(\omega_1+\omega_2)/2}~\tau _A ~ A^c_{-\omega_1}A^c_{-\omega_2} .
\end{equation}

\item \textcolor{red}{$\mathcal{O}=\bar{A}_{-\omega_1}^{c}\bar{A}_{\omega_2}^{c}~~\omega_{1,2}>0$}
\begin{align*}
\langle \{k_\nu, \Delta_\nu \} |  \bar{A}_{-\omega_1}^{c}\bar{A}_{\omega_2}^{c} | \{k'_\nu, \Delta'_\nu \} \rangle & =\langle \Omega^c | \otimes _{\nu \geq 0} ~(A_{\nu} ^c)^{\Delta_\nu + j_\nu}~ (A_{-\nu} ^c)^{k_\nu} ~  \bar{A}_{-\omega_1}^{c}\bar{A}_{\omega_2}^{c} | \{k'_\nu, \Delta'_\nu \} \rangle
\\
 & =\langle \Omega^c | \bar{A}_{-\omega_1}^{c}\otimes _{\nu \geq 0} ~(A_{\nu} ^c)^{\Delta_\nu + j_\nu}~ (A_{-\nu} ^c)^{k_\nu} ~  \bar{A}_{\omega_1}^{c}\bar{A}_{\omega_2}^{c} | \{k'_\nu, \Delta'_\nu \} \rangle
 \\
 &
=e^{-\omega_1/2}\langle \Omega^c |A_{\omega_1}^{c} \otimes _{\nu \geq 0} ~(A_{\nu} ^c)^{\Delta_\nu + j_\nu}~ (A_{-\nu} ^c)^{k_\nu} ~  \bar{A}_{\omega_2}^{c} | \{k'_\nu, \Delta'_\nu \} \rangle
\\
&
=e^{-\omega_1/2}\langle \Omega^c | \otimes _{\nu \geq 0} ~(A_{\nu} ^c)^{\Delta_\nu + j_\nu}~ (A_{-\nu} ^c)^{k_\nu} ~  A_{\omega_1}^{c}\bar{A}_{\omega_2}^{c} | \{k'_\nu, \Delta'_\nu \} \rangle
\\
&
~~+e^{-\omega_1/2}\langle \Omega^c | \otimes _{\nu \geq 0} [A_{\omega_1}^{c},~(A_{\nu} ^c)^{\Delta_\nu + j_\nu}]~ (A_{-\nu} ^c)^{k_\nu} ~  \bar{A}_{\omega_2}^{c} | \{k'_\nu, \Delta'_\nu \} \rangle\\
&
=e^{-\omega_1/2}\langle \{k_\nu, \Delta_\nu \} |  A_{\omega_1}^{c}\bar{A}_{\omega_2}^{c} | \{k'_\nu, \Delta'_\nu \} \rangle
\\& =e^{-\omega_1/2}\langle \{k_\nu, \Delta_\nu \} |  A_{\omega_1}^{c}\bar{A}_{\omega_2}^{c} \otimes _{\nu \geq 0} ~ (A_\nu ^c)^{k'_\nu} ~ (A_{-\nu} ^c)^{\Delta'_\nu + j'_\nu}~| \Omega^c \rangle
\\
& =e^{-\omega_1/2}\langle \{k_\nu, \Delta_\nu \} |  A_{\omega_1}^{c} \otimes _{\nu \geq 0} ~ (A_\nu ^c)^{k'_\nu} ~ (A_{-\nu} ^c)^{\Delta'_\nu + j'_\nu}~\bar{A}_{\omega_2}^{c}| \Omega^c \rangle
\\
&=e^{(\omega_2-\omega_1)/2}\langle \{k_\nu, \Delta_\nu \} |  A_{\omega_1}^{c} \otimes _{\nu \geq 0} ~ (A_\nu ^c)^{k'_\nu} ~ (A_{-\nu} ^c)^{\Delta'_\nu + j'_\nu}~A_{-\omega_2}^{c}| \Omega^c \rangle
\\
&=e^{(\omega_2-\omega_1)/2}\langle \{k_\nu, \Delta_\nu \} |  A_{\omega_1}^{c}A_{-\omega_2}^{c} \otimes _{\nu \geq 0} ~ (A_\nu ^c)^{k'_\nu} ~ (A_{-\nu} ^c)^{\Delta'_\nu + j'_\nu}~| \Omega^c \rangle
\\
& -e^{(\omega_2-\omega_1)/2}\langle \{k_\nu, \Delta_\nu \} |  A_{\omega_1}^{c} \otimes _{\nu \geq 0} [A_{-\omega_2}^{c},~ (A_\nu ^c)^{k'_\nu}] ~ (A_{-\nu} ^c)^{\Delta'_\nu + j'_\nu}~| \Omega^c \rangle
\\
& =e^{(\omega_2-\omega_1)/2}\langle \{k_\nu, \Delta_\nu \} |  A_{\omega_1}^{c}A_{-\omega_2}^{c} \otimes _{\nu \geq 0} ~ (A_\nu ^c)^{k'_\nu} ~ (A_{-\nu} ^c)^{\Delta'_\nu + j'_\nu}~| \Omega^c \rangle
\\
& =e^{(\omega_2-\omega_1)/2}\langle \{k_\nu, \Delta_\nu \} |  A_{\omega_1}^{c}A_{-\omega_2}^{c} | \{k'_\nu, \Delta'_\nu \} \rangle
\end{align*}
and we get
\begin{equation}
    \mathcal{F}(\bar{A}^c_{\omega_1} \bar{A}^c_{\omega_2}) = e^{(\omega_2-\omega_1)/2}~\tau _A ~ A^c_{\omega_1}A^c_{-\omega_2} .
\end{equation}

\item \textcolor{red}{$\mathcal{O}=\bar{A}_{\omega_1}^{c}\bar{A}_{-\omega_2}^{c}~~\omega_{1,2}>0$}
\begin{align*}
    \langle \{k_\nu, \Delta_\nu \} |  \bar{A}_{\omega_1}^{c}\bar{A}_{-\omega_2}^{c} | \{k'_\nu, \Delta'_\nu \} \rangle=e^{(\omega_1-\omega_2)/2}\langle \{k_\nu, \Delta_\nu \} |  A_{-\omega_1}^{c}A_{\omega_2}^{c} | \{k'_\nu, \Delta'_\nu \} \rangle
\end{align*}
which gives
\begin{equation}
    \mathcal{F}(\bar{A}^c_{\omega_1} \bar{A}^c_{-\omega_2}) = e^{(\omega_1-\omega_2)/2}A^c_{-\omega_1}A^c_{\omega_2} ~\tau _A.
\end{equation}

\item \textcolor{red}{$\mathcal{O}=\bar{A}_{-\omega_1}^{c}\bar{A}_{-\omega_2}^{c}~~\omega_{1,2}>0$}
\begin{align*}
    \langle \{k_\nu, \Delta_\nu \} |  \bar{A}_{-\omega_1}^{c}\bar{A}_{-\omega_2}^{c} | \{k'_\nu, \Delta'_\nu \} \rangle=e^{-(\omega_1+\omega_2)/2}\langle \{k_\nu, \Delta_\nu \} |  A_{\omega_1}^{c}A_{\omega_2}^{c} | \{k'_\nu, \Delta'_\nu \} \rangle
\end{align*}
which gives
\begin{equation}
    \mathcal{F}(\bar{A}^c_{-\omega_1} \bar{A}^c_{-\omega_2}) = e^{-(\omega_1+\omega_2)/2}A^c_{-\omega_1}A^c_{-\omega_2} ~\tau _A.
\end{equation}

\end{itemize}


\section{Warmup: leading term reconstruction through the Petz map}

As it is explained in Sec. \ref{petz} to leading order, the entanglement wedge reconstruction can be done through the Petz map.  It has been discussed in more detail in \cite{Bahiru:2022ukn}, and here we will extend the discussion briefly in the context of the projected Hamiltonian. The Petz map is given as
\begin{align} \label{petzmap}
    \Phi_A(X) =& \Tr _{\Bar{A}} [  P_{code}'] ^{-1/2}~~  \Tr _{\Bar{A}} [  P_{code}'~ \Phi_{HKLL}(X)~ P_{code}']  ~~ \Tr _{\Bar{A}} [  P_{code}'] ^{-1/2}
    \\
    =& \int dt d\Omega~ K(X|t,\Omega)~ \tau_A^{-1/2} ~ \Tr _{\Bar{A}} [  P_{code}'~P_0~ O(t,\Omega)~P_0~ P_{code}']~ \tau_A^{-1/2}
    \\
    =& \int dt d\Omega~ K(X|t,\Omega)~ \tau_A^{-1/2} ~ \Tr _{\Bar{A}} [  P_{code}'~e^{i\hat{H}^{code} t}~ O^c(\Omega)~ e^{-i\hat{H}^{code} t}~ P_{code}']~ \tau_A^{-1/2}
\end{align}
where $  \tau_A =\Tr _{\Bar{A}} [  P_{code}'] $.

Since at the leading term, we deal with free fields in the bulk, which are dual to the GFF sector over the vacuum of the CFT, we have 
\begin{equation}
    \mathcal{H}_{code} = \mathcal{H}_{GFF},~~~~~~~~~~~ P_{code} =P_{GFF}.
\end{equation}
Therefore, 
\begin{equation}
   O^c(t=0, \Omega) = O_0^c(\Omega) = \sum_{nlm} g_{nlm}( \Omega) O_{nlm} + h.c.
\end{equation}
where we denote the GFF by $O_0^c(t, \Omega)$ and 
\begin{equation}
    g_{nlm}(\Omega) = \lim_{r\rightarrow \infty} r^{\Delta} f_{nlm} (r, t=0, \Omega)
\end{equation}
and as we mentioned from the extrapolate dictionary, we arrive to
\begin{equation}
    V ~ a_{nlm}~ V^\dagger  = O_{nlm},~~~~~~~~~V ~ a^\dagger_{nlm}~ V^\dagger = O^\dagger_{nlm}.
\end{equation}
while $ [O_{nlm}, O^\dagger_{n'l'm'}] = \delta_{nn'} \delta_{ll'} \delta_{mm'}$. Moreover, from the constraints of general relativity  in the canonical quantization of gravity, we have 
\begin{equation}\label{gffhamil}
    \hat{H}_{0} = \hat{H}^{code} - \alpha I = \sum _{nlm } \omega_{nlm} O_{nlm}^\dagger O_{nlm}:= H_{GFF} .
\end{equation}
Therefore
\begin{equation}\label{primecode}
    O_0^c(t, \Omega) =  e^{i \hat{H}^{code} t} ~ O_0^c(t=0,\Omega)~e^{-i \hat{H}^{code }t} = \sum _{nlm} \Tilde{g}_{nlm}(t, \Omega) O_{nlm} +h.c.
\end{equation}
where 
\begin{equation}\label{tildeg}
    \Tilde{g}_{nlm}(t, \Omega) := g_{nlm}( \Omega) e^{-i \omega_{nlm} t}.
\end{equation}
We use the fact that $\lambda \sim O(1)$, which means the interactions between matters is much stronger than gravitational interaction. In case that  $\lambda \ll 1$, we need to consider also the Hilbert space of the gravitational part which we should deal perturbatively, as an example look at \cite{Chowdhury:2021nxw}.

We note that at the leading term in the large $N$ expansion,  we deal with quantum field theory on curved spacetime, and thus we have a non-dynamical theory of gravity, and the gravitational part does not contribute to the ADM Hamiltonian. 
Moreover, it is needed to mention that $ H_{GFF}$ is only well-defined in the code subspace since the GFF are scalars of the boundary theory with conformal dimension $ \Delta = d/2 + \sqrt{m^2 + d^2/4} $, and it's actually not a real scalar since in a CFT in $d$- spacetime dimension, the condition that a scalar operator is free is equivalent to the fact that $ \Delta = d/2 -1$. 
The GFF sector does not obey the linear equation of motion in the CFT, and we can not describe them in terms of a local free Lagrangian in the spacetime background in which the CFT lives.

By substituting \eqref{primecode} in \eqref{petzmap}, we find that the Petz map reconstruction of an operator in the entanglement wedge is given by 
\begin{equation}
\begin{aligned}
        \Phi_A(X) = \sum _{nlm} F_{nlm}(X)  ~ \tau_A^{-1/2} ~ &\Tr _{\Bar{A}} [  P_{code}'~ O_{nlm}~  P_{code}']~ \tau_A^{-1/2}
        \\
        & + F^*_{nlm}(X)  ~ \tau_A^{-1/2} ~ \Tr _{\Bar{A}} [  P_{code}'~ O^\dagger_{nlm}~  P_{code}']~ \tau_A^{-1/2} 
\end{aligned}
\end{equation}
where 
\begin{equation}
    F_{nlm}(X) = \int dt d\Omega~ K(X|t,\Omega)~ \Tilde{g}_{nlm}(t, \Omega) = f_{nlm}^{0}(X)
\end{equation}
When the backreaction is big enough and the semi-classical geometry is not pure AdS anymore, we do not have the equality above between $F_{nlm}(X)$ and $  f_{nlm}^{0}(X)$ since we have the modification in the smearing function of the even first term of the global HKLL reconstruction.

In order to proceed, let us choose the set of eigenfunctions of the projected modular Hamiltonians of the corresponding regions, $ \{A _\nu\}$, and $ \{ \Bar{A}_\nu\}$, respectively, as a basis for the operator algebra of the regions projected to the code subspace.
First, we need to write the global modes $O_{nlm}$ as a linear combination of these two sets of eigenfunctions as
\begin{equation}\label{primary}
    O_{nlm} = \sum_{\nu} \alpha^A_{nlm,\nu}~ A_\nu + \alpha^{\Bar{A}}_{nlm,\nu}~ \bar{A}_\nu 
\end{equation}
where the coefficients $\alpha^A_{nlm,\nu} $, and $ \alpha^{\Bar{A}}_{nlm,\nu}$ are obtained from the Bogoliubov transformation  between the global modes $ a_{nlm}$ and eigenfunctions of the modular Hamiltonian of the regions $ a = \mathcal{E}_A$ and $ \bar{a} = \mathcal{E}_{\bar{A}}$ as a result of the JLMS statement \cite{Bahiru:2022ukn}.

The Petz map reconstruction of the bulk field $\phi(X)$ arrives at 
\begin{equation}
 \Phi_A(X) = \sum _{\nu} \mathcal{F}^A_{\nu}(X)  ~ \tau_A^{-1/2} ~ \Tr _{\Bar{A}} [  P_{code}'~ A_{\nu}~  P_{code}']~ \tau_A^{-1/2} + \mathcal{F}^{\bar{A}*}_{\nu}(X)  ~ \tau_A^{-1/2} ~ \Tr _{\Bar{A}} [  P_{code}'~ \bar{A}_{\nu}~  P_{code}']~ \tau_A^{-1/2} 
\end{equation}
where
\begin{equation}
    \mathcal{F}^I_{\nu}(X) = \sum_{nlm} F_{nlm}(X) ~\alpha^I_{nlm,\nu} + F_{nlm}^*(X)~ \alpha^{I*}_{nlm,\nu}
\end{equation}
for $ I \in \{A, \bar{A}\}$.
Moreover, one can find that by the definition of the Bogoliubov transformation, we have 
\begin{equation}
    \mathcal{F}^{\bar{A}}_{\nu}(X) = \sum _{nlm}   f_{nlm}(X) ~\alpha^{\bar{A}}_{nlm,\nu} + f_{nlm}^*(X)~ \alpha^{\bar{A}*}_{nlm,-\nu}=0, ~~~~~~~~\forall X \in a.
\end{equation}
and therefore, the Petz map reconstruction reads off as 
\begin{equation}
     \Phi_A(X) = \sum _{\nu} \mathcal{F}^A_{\nu}(X) ~ \tau_A^{-1/2} ~ \Tr _{\Bar{A}} [  P_{code}'~ A_{\nu}~  P_{code}']~ \tau_A^{-1/2}.
\end{equation}
By using the equality \eqref{111}, and the fact that for an operator $O_A$ have only support on region $A$ and commute with every operator $X_{\bar{A}}$, one can show that $\tau_A^{-1}$ commute with the $ \tr_{\bar{A}}\big( P_{code} O_A P_{code}\big)$, we arrive at 
\begin{equation}
     \Phi_A(X) = \sum _{\nu} \mathcal{F}^A_{\nu}(X)~ A_{\nu}~  
\end{equation}
using the fact that 
\begin{equation}
     \mathcal{F}^A_{\nu}(X) = \int _{bdy} dt d\Omega~ K(X|t, \Omega) ~ \sum_{nlm} \big( \tilde{g}_{nlm}(t,\Omega)~ \alpha^A _{n, \nu} +  \tilde{g}^*_{nlm}(t,\Omega)~ \alpha^{A*} _{n, -\nu} \big)
\end{equation}
and the fact that $ A_\nu$ in the free theory ( bulk side) is given by two labels $ (\omega, X_s)$ where $ X_s$ is a co-dimension 2 surface  $ S $ \cite{Faulkner:2017}
\begin{equation}
    \Phi_\omega(X_s) = \int ds~ e^{i s \omega}~ e^{i K_a s } ~\phi(X_s)~ e^{-i K_a s}, ~~~~ \forall X_s \in S.
\end{equation}
Consider $S$ to be the region $A$ itself, and extrapolate dictionary, we have the final formula for the Petz map reconstruction at the large $N$ limit as 
\begin{equation}
    \Phi_A(X) = \int_{-\infty}^{\infty} ds \int _{y_A} dy ~ K_{Petz, A}(X|s, y_A)~  e^{i K^c_A s } ~O^c(y_A)~ e^{-i K^c_A s}
\end{equation}
where the smearing function is given by 
\begin{equation}
    K_{Petz, A}(X|s, y_A) = \sum _{nlm} \int d\omega ~ e^{i s \omega}~ \big( F_{nlm}(X) ~\alpha^{A}_{nlm}(\omega, y_A) +  F^*_{nlm}(X) ~\alpha^{A,*}_{nlm}(-\omega, y_A)  \big)
\end{equation}
where $\alpha^{A}_{nlm}(\omega, y_A) $ are the Bogoliubov coefficients between $ O_{nlm}$ and $O_\omega(y_A)$.  In particular, in \cite{Bahiru:2022ukn} it has been shown that in the case of the AdS-Rindler wedge, we get exactly the same smearing distribution as the causal wedge HKLL reconstruction.

Here, we have the reconstruction of the bulk operator in the entanglement wedge from the boundary, which has its support only on the region $A$, but it is restricted to the code subspace, and it has a correct action on the code subspace by construction. 
But, Entanglement wedge reconstruction is  the statement that says any bulk operator $ \phi_a$ acting within $\mathcal{H}_a$ can be represented in the CFT with an operator $\Phi_A$
that has support only on $\mathcal{H}_A$.
As a result, by considering the decomposition of the CFT Hilbert space as 
\begin{equation}
    \mathcal{H}_{CFT} = \mathcal{H}_{code} \oplus \mathcal{H}_{code} ^\perp,
\end{equation}
one can choose any operator $O^\perp \in  \mathcal{L}(\mathcal{H}_{code} ^\perp) \cap \mathcal{L}(\mathcal{H}_A) $ and then 
$\Phi_A(X) \oplus O^\perp $ is a CFT reconstruction of the bulk field and in particular 
\begin{equation}
    \Phi_A(X) \oplus O^\perp  
\end{equation}
is a reconstruction of the bulk operator in the region $A$.

One natural extension of the operator to the entire $\mathcal{H}_A$ is to drop $P_{code}$ and, as a final result, one specific choice of the entanglement wedge reconstruction is 
\begin{equation}\label{finalewrfree}
    \Phi_A(X) = \int_{-\infty}^{\infty} ds \int _{y_A} dy ~ K_{Petz, A}(X|s, y_A)~  e^{i K_A s } ~ O(y_A)~ e^{-i K_A s} 
\end{equation}
Here we reproduce the same result as \cite{Faulkner:2017} where they used the modular flow approach. 

We note again that the action of the CFT reconstructions on the code subspace should be unique, but in general, a bulk field can have a lot of different boundary representations, which reflects the ability of the code to correct for a variety of erasures.


\section{EWR up to 1/N correction through the twirled Petz map}

In order to do the entanglement wedge reconstruction, we should use the twirled Petz map as
\begin{align}\label{twiled}
    \Phi_A(X) = & \int _{\mathbb{R}} ds \beta_0(s) 
    \Tr _{\Bar{A}} [  P_{code}'] ^{-(1+is)/2}~~  \Tr _{\Bar{A}} [  P_{code}'~ \Phi_{HKLL}(X)~ P_{code}']  ~~ \Tr _{\Bar{A}} [  P_{code}'] ^{-(1-is)/2}
    \\
    =& \int _{\mathbb{R}} ds \beta_0(s)  \int dt d\Omega~ K(X|t,\Omega)~
    \\
    &\tau_A^{-(1+is)/2} ~ \Tr _{\Bar{A}} [  P_{code}'~P_0~ O(t,\Omega)~P_0~ P_{code}']~ \tau_A^{-(1-is)/2}
    \\
    +& \frac{\lambda}{N} \int _{\mathbb{R}} ds \beta_0(s)  \int ~ dX'~ dt_1 d\Omega_1~ dt_2 d\Omega_2~G(X|X')~  K(X'|t_1,\Omega_1) K(X'|t_2,\Omega_2)
    \\
    &  \tau_A^{-(1+is)/2} ~ \Tr _{\Bar{A}} [  P_{code}'~P_0~ O(t_1,\Omega_1)~O(t_2,\Omega_2)~P_0~ P_{code}']~ \tau_A^{-(1-is)/2}
\end{align}
In this case, since the bulk fields are not free, the code subspace is not the GFF sector anymore. Thus, how can we treat the problem? 

One can use the fact that in quantum field theory, in an interacting theory where the Hamiltonian $H$ differs from the free Hamiltonian $H_0$, the Heisenberg equation of motion is  still satisfied, and we can write the interacting field as 
\begin{equation}
    \phi(r,t, \Omega) = \sum_{nlm}\Big( f_{nlm}(r,t,\Omega) a_{nlm}(t)  + h.c.\Big)
\end{equation}
$ a_{nlm}(t)$ and $ a_{nlm}^\dagger(t)$ are the interacting creation and annihilation operators in the theory of any fixed time $t$ and we have $ [ a_{nlm}(t),  a_{n'l'm'}^\dagger(t)] = \delta_{nn'}\delta_{ll'}\delta_{mm'}$. Therefore, the Fock space is the same at every time due to the time-translational invariance of the AdS spacetime. 

Finally, we can define the interacting modes to be equal to free modes at any fixed time $ t=t_0 $ as $  a_{nlm}(t = t_0) :=  a_{nlm}$ and as a result
\begin{equation}
    \phi(r,t_0, \Omega) = \phi_0(r, t_0, \Omega)
\end{equation}
where $\phi_0(r, t, \Omega) $ is the free field on AdS background.

The Hamiltonian of the interacting theory is given by 
\begin{equation}
    H_{bulk,matter} = H_0 +H' = H_0 + \frac{\lambda}{N}\int dr d\Omega~ \sqrt{-g}~ \frac{1}{3} \phi^3(t=0, r, \Omega)
\end{equation}
We can put the modes of the interacting theory and free theory at $t_0 =0$ to be equal and then we have 
\begin{equation}
     \phi(r,t_0=0, \Omega) = \phi_0(r, t_0=0, \Omega).
\end{equation}

If we assume that we have a small amount of backreaction, we will get 
\begin{equation}
    \hat{H} = \hat{H}^{code}-\alpha I = V~ (H_0 +H') ~V^\dagger = H_{GFF} + \frac{\lambda}{N}\int dr d\Omega~ \frac{\sqrt{-g}}{3} V\phi_0^3(t=0, r, \Omega)V^\dagger 
\end{equation}
and to find the exact form of the $ \hat{H}_{ADM}$ one can use the relation \eqref{mapmodes} and to first correction in $1/N$ we have 
\begin{equation}
     \hat{H} = \hat{H}^{code}-\alpha I = H_{GFF} + \frac{\lambda}{N} ~H'
\end{equation}
where 
\begin{equation}
    H' = \int dr d\Omega~ \frac{\sqrt{-g}}{3}  \Phi_{HKLL,0}^{c3}(t=0, r, \Omega)
\end{equation}
Now, to find the twirled Petz map reconstruction we should deal with the two terms below
\begin{itemize}
    \item First term
\begin{equation}
    \Tr _{\Bar{A}} [  P_{code}'~P_0~ O(t,\Omega)~P_0~ P_{code}'] = \Tr _{\Bar{A}} [  P_{code}'~e^{i \hat{H}_{ADM}t}~ O_0^c(t=0,\Omega)~e^{-i \hat{H}_{ADM}t}~ P_{code}']
\end{equation}

    \item  Second term

    \begin{align}
   (1/N)\Tr _{\Bar{A}} [ & P_{code}'~P_0~ O(t_1,\Omega_1)~  O(t_2,\Omega_2) ~P_0~ P_{code}'] =\\
   &  (1/N)  \Tr _{\Bar{A}} [  P_{code}'~P_{GFF}~ O^c_0(t_1,\Omega_1)~O^c_0(t_2,\Omega_2)~P_{GFF}~ P_{code}'] + O(1/N^2)
    \end{align}
\end{itemize}

We introduce 
\begin{equation}
    O^c_0(t,\Omega) = e^{i \hat{H}_{GFF}t}~ O_0^c(t=0,\Omega)~e^{-i \hat{H}_{GFF}t}= \sum _{nlm} \Tilde{g}_{nlm}(t, \Omega) O_{nlm} +h.c..
\end{equation}
while $  \Tilde{g}_{nlm}(t, \Omega)$ is given by the relation \eqref{tildeg}.

\subsection{Expansion of the projected HKLL operators in terms of GFF modes}

The projected HKLL operators up to $1/N$ correction on $ \mathcal{H}_{code}=\mathcal{H}_0$ in terms of the GFF modes can be obtained as 

\begin{equation}
  \begin{aligned}
    \Phi_{HKLL}^{c} (X)= & P_0 ~\Phi_{HKLL}^{(0)}(X)~ P_0  + \frac{1}{N} P_0 ~\Phi_{HKLL}^{(1)}(X)~ P_0 + O(1/N^2)
    \\
   = &\sum_{nlm}~  \Big( f_{nlm}( X) O_{nlm} +  f_{nlm}^*( X) O^\dagger_{nlm} \Big)
   \\
   + &  \frac{\lambda}{N} \sum_{nlm} \sum_{n'l'm'} ~ \Big( \mathcal{H}^{(1)}_{(nlm);(n'l'm')}  (X) ~ O_{nlm} O_{n'l'm'}+ \mathcal{H}^{(2)}_{(nlm);(n'l'm')} (X) ~ O_{nlm} O^\dagger_{n'l'm'}
   \\
   &~~~~~~~~~~~~~~~~~~~~~~ + \mathcal{H}^{(2)*}_{(nlm);(n'l'm')} (X) ~ O^\dagger_{nlm} O_{n'l'm'}+ \mathcal{H}^{(1)*}_{(nlm);(n'l'm')} (X) ~ O^\dagger_{nlm} O^\dagger_{n'l'm'} \Big)
\end{aligned}  
\end{equation}
where 
\begin{equation}
    \mathcal{H}^{(z)}_{(nlm);(n'l'm')}(X) = \mathcal{K}^{(z)}_{(nlm);(n'l'm')} +i~ \mathcal{I}^{(z)}_{(nlm);(n'l'm')}, ~~~~~~ \forall z \in \{1,2\}
\end{equation}
and $ \mathcal{I}^{(z)}_{(nlm);(n'l'm')}$, $\mathcal{K}^{(z)}_{(nlm);(n'l'm')}$ are given by \eqref{I} and \eqref{K} respectively. The details are provided below.

\subsubsection{First Term}

In order to find the contribution of the  first term in the twirled Petz map reconstruction, we need to evaluate 
\begin{equation}
     O^c(t,\Omega) =e^{i \hat{H}^{code}t}~ O_0^c(t=0,\Omega)~e^{-i \hat{H}^{code}t}= e^{i \hat{H}t}~ O_0^c(t=0,\Omega)~e^{-i \hat{H} t}
\end{equation}
To the first correction in $1/N$ perturbation theory, using the machinery of the interacting picture, we have
\begin{equation}
    \exp \Big( i \hat{H} (t-t_0) \Big) =  e^{i \hat{H}_{GFF}t}~ \Big( I + i \frac{\lambda}{N} \int_{t_0}^t dt'~ H_I'(t')\Big)~e^{-i \hat{H}_{GFF}t_0}  + O (1/N^2)
\end{equation}
where
\begin{equation}
\begin{aligned}
         H_I'(t) = & e^{i \hat{H}_{GFF}t}~ H'(t)~e^{-i \hat{H}_{GFF}t} =  O_0^c(t, \Omega)^3 
         \\
         =& e^{i \hat{H}_{GFF}t} \int dr d\Omega~ \frac{\sqrt{-g}}{3}(\Phi^{(0),c}_{HKLL}(s =0 , r, \Omega))^3 ~ e^{- i \hat{H}_{GFF}t}
         \\
         =&  \int dr d\Omega~ \frac{\sqrt{-g}}{3} ~(\Phi^{(0),c}_{HKLL}(s, r, \Omega))^3 
\end{aligned}
\end{equation}
where
\begin{equation}
    \Phi^{(0),c}_{HKLL}(s, r, \Omega) = \int dt' d\Omega' ~K(s,r, \Omega| t',\Omega')~ O^c_0(t', \Omega')
\end{equation}
Therefore, we reach 
\begin{equation}
    O^c(t,\Omega)= O_0^c(t,\Omega) + i \frac{\lambda}{N}~ e^{i \hat{H}_{GFF}t}\Big( \int_0^t dt'~ [H_I'(t'),  O_0^c(t=0,\Omega)]\Big)e^{-i \hat{H}_{GFF}t} + O(1/N^2)
\end{equation}

Consider the fact that one can expand $ O_0^c(t, \Omega)^3$ in terms of GFF modes as 
\begin{equation}
    O_0^c(t, \Omega)^3 = \sum_{nlm} \sum_{n'l'm'}\sum_{n"l"m"} ~\tilde{g}_{nlm}(t, \Omega) \tilde{g}_{n'l'm'}(t, \Omega)\tilde{g}_{n"l"m"}(t, \Omega) ~ O_{nlm} O_{n'l'm'}O_{n"l"m"} + ...
\end{equation}
as a result, we obtain 
\begin{equation}
  \begin{aligned}
    \Phi_{HKLL}^{(0),c} (X)= & P_0 ~\Phi_{HKLL}^{(0)}(X)~ P_0 
    \\
   = &\sum_{nlm}~  \Big( f_{nlm}( X) O_{nlm} +  f_{nlm}^*( X) O^\dagger_{nlm} \Big)
   \\
   + & i \frac{\lambda}{N} \sum_{nlm} \sum_{n'l'm'} ~ \Big( \mathcal{I}^{(1)}_{(nlm);(n'l'm')}  (X) ~ O_{nlm} O_{n'l'm'}+ \mathcal{I}^{(2)}_{(nlm);(n'l'm')} (X) ~ O_{nlm} O^\dagger_{n'l'm'}
   \\
   &~~~~~~~~~~~~~~~~~~~~~~ + \mathcal{I}^{(2)*}_{(nlm);(n'l'm')} (X) ~ O^\dagger_{nlm} O_{n'l'm'}+ \mathcal{I}^{(1)*}_{(nlm);(n'l'm')} (X) ~ O^\dagger_{nlm} O^\dagger_{n'l'm'} \Big)
\end{aligned}  
\end{equation}
where 
\begin{equation}
    f_{nlm}(X)
  = \int d\Omega'\, dt'\, K(X|t', \Omega)\, \tilde{g}_{nlm}(t', \Omega')
\end{equation}
is the bulk mode functions and the 0th-order interaction kernels are
\begin{equation}\label{I}
 \begin{aligned}
\mathcal{I}^{(1)}_{(nlm);(n'l'm')} &(X) =
\int _{y=(t, \Omega)} dt d\Omega~ K(X|t,\Omega)~\int_0^{t} dt'  \int dr' d\Omega' _{X'=(r',t',\Omega')}~\sqrt{-g} 
  \\[2mm]
  &f_{nlm}(X') f_{n'l'm'}(X') ~e^{-it\omega_{nlm}} ~ e^{-it\omega_{n'l'm'}}\, \sum_{n"l"m"}\Big[f_{n"l"m"}(X)g_{n"l"m"}^*(\Omega) - f_{n"l"m"}^*(X)g_{n"l"m"}(\Omega)\Big]
 ,
\\[2mm]
\mathcal{I}^{(2)}_{(nlm);(n'l'm')} &(X) =
\int _{y=(t, \Omega)} dt d\Omega~K(X|t,\Omega)~ \int_0^{t} dt'  \int dr' d\Omega' _{X'=(r',t',\Omega')}~\sqrt{-g}~ 
  \\[2mm]
  &f_{nlm}(X') f^*_{n'l'm'}(X')e^{-it\omega_{nlm}} ~ e^{+it\omega_{n'l'm'}} \, \sum_{n"l"m"}\Big[f_{n"l"m"}(X)g_{n"l"m"}^*(\Omega) - f_{n"l"m"}^*(X)g_{n"l"m"}(\Omega)\Big].
\end{aligned}   
\end{equation}


\subsubsection{Second Term}

Since we are only keeping the terms up to the first correction in $1/N$, we can evolve the $\mathcal{O}^c_0$ in the second term using only the leading term of the code subspace Hamiltonian, which is the  GFF Hamiltonian we introduced in \eqref{gffhamil} in the code subspace of the boundary theory. Therefore, we need to find 
\begin{equation}
     (1/N)~  O^c(t_1,\Omega_1)~O^c(t_2,\Omega_2)= (1/N)~  O^c_0(t_1,\Omega_1)~O^c_0(t_2,\Omega_2) + O( 1/N^2)
\end{equation}
One can find that 
\begin{equation}
\begin{aligned}
       O^c_0(t_1,\Omega_1)~O^c_0(t_2,\Omega_2) = &
       \\
       \sum_{nlm} \sum_{n'l'm'} ~ \Big( \tilde{g}_{nlm} &(t_1, \Omega_1) \tilde{g}_{n'l'm'}(t_2, \Omega_2)~ O_{nlm}O_{n'l'm'} + \tilde{g}_{nlm}(t_1, \Omega_1) \tilde{g}^*_{n'l'm'}(t_2, \Omega_2)~ O_{nlm}O^\dagger_{n'l'm'} +h.c. \Big)
\end{aligned}
\end{equation}
Therefore, we arrive to 
\begin{equation}
 \begin{aligned}
    \Phi_{HKLL}^{(1),c} (X)= & P_0 ~\Phi_{HKLL}^{(1)}(X)~ P_0 
    \\
   = & \lambda \sum_{nlm} \sum_{n'l'm'} ~ \Big( \mathcal{K}^{(1)}_{(nlm);(n'l'm')}  (X) ~ O_{nlm} O_{n'l'm'}+ \mathcal{K}^{(2)}_{(nlm);(n'l'm')} (X) ~ O_{nlm} O^\dagger_{n'l'm'}
   \\
   &~~~~~~~~~~~~~~~~~~~~~~ + \mathcal{K}^{(2)*}_{(nlm);(n'l'm')} (X) ~ O^\dagger_{nlm} O_{n'l'm'}+ \mathcal{K}^{(1)*}_{(nlm);(n'l'm')} (X) ~ O^\dagger_{nlm} O^\dagger_{n'l'm'} \Big)
\end{aligned}   
\end{equation}
where the 1st-order interaction kernels are
\begin{equation}\label{K}
    \begin{aligned}
       \mathcal{K}^{(1)}_{(nlm);(n'l'm')} & = \int dX' \sqrt{-g} ~G(X|X') ~ f_{nlm}(X') ~f_{n'l'm'}(X')
       \\
       \mathcal{K}^{(2)}_{(nlm);(n'l'm')}&= \int dX' \sqrt{-g} ~G(X|X') ~ f_{nlm}(X') ~f^*_{n'l'm'}(X')
    \end{aligned}
\end{equation}

\subsection{Expansion of the projected HKLL operators in terms of the eigenbasis of the modular Hamiltonians}

Considering the eigenfunctions of modular Hamiltonians $K_A^c$ and $K^c_{\bar{A}}$
$\{ A^{c,A}_\omega = A^c_\omega\}$ and $\{ A^{c,\bar{A}}_\omega = \bar{A}^c_\omega\}$
as the basis for the operator algebra of the corresponding regions restricted to the code subspace. We have
\begin{equation}
    \begin{aligned}
        O_{nlm}& = \sum_\omega ~\alpha^A_{nlm;\omega}~ A_\omega + \alpha^{\bar{A}}_{nlm;\omega}~ \bar{A}_\omega
        \\
        O^\dagger_{nlm}& = \sum_\omega ~\alpha^{A*}_{nlm;-\omega}~ A_\omega + \alpha^{\bar{A}*}_{nlm;-\omega}~ \bar{A}_\omega
    \end{aligned}
\end{equation}
where the Bogoliubov coefficients are the same as the bulk Bogoliubov coefficients of decomposing $a_{nlm}$ in terms of $\{ A^{c,A}_\omega = A^c_\omega\}$ and $\{ A^{c,\bar{A}}_\omega = \bar{A}^c_\omega\}$, i.e. 
\begin{equation}
    \alpha^A_{nlm;\omega} = \alpha^a_{nlm;\omega}, ~~~~~~~~~~~ \alpha^{\bar{A}}_{nlm;\omega}= \alpha^{\bar{a}}_{nlm;\omega}.
\end{equation}

\subsubsection{Leading term}
The leading term can be rewritten as 
\begin{equation}\label{leadterm}
    \sum_{nlm}~  \Big( f_{nlm}( X) O_{nlm} +  f_{nlm}^*( X) O^\dagger_{nlm} \Big) = \sum _{I \in \{A, \bar{A}\}} \sum_\omega ~ \mathcal{F}^I_\omega (X)~ A_\omega ^{c,I}
\end{equation}
where 
\begin{equation}
    \mathcal{F}^I_\omega (X) = \sum _{nlm} \Big( f_{nlm}(X) ~\alpha^I_{nlm;\omega} + f^*_{nlm}(X) ~\alpha^{I*}_{nlm;-\omega} \Big)
\end{equation}
Here, we should be careful about the constraints coming from the relation between the Bogoliubov coefficients. 

The commutation relation between the eigenfunctions of the modular Hamiltonians leads to 
\begin{equation}
    \sum_{I\in \{A, \bar{A}\}} \sum_{\nu, \nu'} \Big( \alpha^I _{nlm;\nu} ~\alpha^{I*}_{n'l'm';\nu'} - \alpha^{I*} _{nlm;-\nu} ~\alpha^{I}_{n'l'm';-\nu'}\Big) = \delta_{nn'} \delta_{ll'} \delta_{mm'}
\end{equation}
and from substituting \eqref{leadterm} in the global GFF expansion of the projected primaries \eqref{primary}, one can finds that 
\begin{equation}
    \begin{aligned}
         & \sum _{nlm} \Big( f_{nlm}(X) ~\alpha^A_{nlm;\omega} + f^*_{nlm}(X) ~\alpha^{A*}_{nlm;-\omega} \Big) = \mathcal{F}^A_\omega (X) =0, ~~~~~~~~~\forall X \in \bar{a},
          \\
         & \sum _{nlm} \Big( f_{nlm}(X) ~\alpha^{\bar{A}}_{nlm;\omega} + f^*_{nlm}(X) ~\alpha^{\bar{A}*}_{nlm;-\omega} \Big) = \mathcal{F}^{\bar{A}}_\omega (X) =0, ~~~~~~~~~\forall X \in a.
    \end{aligned}
\end{equation}
Therefore, since we are interested in the EWR of the bulk field in the bulk region $ X \in a = \mathcal{E}_A$, we have 
\begin{equation}\label{leading}
       \sum_{nlm}~  \Big( f_{nlm}( X) O_{nlm} +  f_{nlm}^*( X) O^\dagger_{nlm} \Big) =  \sum_\omega ~ \mathcal{F}^A_\omega (X)~ A_\omega ^{c,A}, ~~~~~~~\forall X \in a.
\end{equation}

\subsubsection{First subleading term}

The subleading term can be rewritten as
\begin{equation}
    \begin{aligned}\label{subleading}
          \frac{1}{N} \Phi_{HKLL}^{c (1)} (X)= 
& \frac{\lambda}{N} \sum_{nlm} \sum_{n'l'm'} ~ \Big( \mathcal{H}^{(1)}_{(nlm);(n'l'm')}  (X) ~ O_{nlm} O_{n'l'm'}+ \mathcal{H}^{(2)}_{(nlm);(n'l'm')} (X) ~ O_{nlm} O^\dagger_{n'l'm'}
   \\
   &~~~~~~~~~~~~~~~~~~~~~~ + \mathcal{H}^{(2)*}_{(nlm);(n'l'm')} (X) ~ O^\dagger_{nlm} O_{n'l'm'}+ \mathcal{H}^{(1)*}_{(nlm);(n'l'm')} (X) ~ O^\dagger_{nlm} O^\dagger_{n'l'm'} \Big) 
   \\
   =& \frac{\lambda}{N} \sum_{\omega_1, \omega_2} \Big(\mathcal{M}^{(1)}_{\omega_1, \omega_2}(X)~ A_{\omega_1} A_{\omega_2} + 
   \mathcal{M}^{(2)}_{\omega_1, \omega_2}(X)~ A_{\omega_1} \bar{A}_{\omega_2}+
   \mathcal{M}^{(3)}_{\omega_1, \omega_2}(X)~ \bar{A}_{\omega_1} \bar{A}_{\omega_2}\Big)
    \end{aligned}
\end{equation}
where 
\begin{equation}
    \begin{aligned}
        \mathcal{M}^{(1)}_{\omega_1, \omega_2}(X) =& \sum_{nlm;n'l'm'} \Big(\mathcal{H}^{(1)}_{(nlm);(n'l'm')}  (X) ~\alpha_{nlm;\omega_1}^A~\alpha_{n'l'm';\omega_2}^A            + \mathcal{H}^{(2)}_{(nlm);(n'l'm')}  (X) ~\alpha_{nlm;\omega_1}^{A} ~\alpha_{n'l'm';-\omega_2}^{A*}
        \\
       & ~~~ + \mathcal{H}^{(2)*}_{(nlm);(n'l'm')}  (X) ~\alpha_{nlm;-\omega_2}^{A*} ~\alpha_{n'l'm';-\omega_1}^{A} +  + \mathcal{H}^{(1)*}_{(nlm);(n'l'm')}  (X) ~\alpha_{nlm;-\omega_2}^{A*} ~\alpha_{n'l'm';-\omega_1}^{A*} \Big)
        \\
        \mathcal{M}^{(2)}_{\omega_1, \omega_2}(X)=&
        \sum_{nlm;n'l'm'}\Big(
        \mathcal{H}^{(1)}_{(nlm);(n'l'm')} (X)
        \Big( \alpha_{nlm;\omega_1}^{A} ~\alpha_{n'l'm';\omega_2}^{\bar{A}} + \alpha_{nlm;\omega_2}^{\bar{A}} ~\alpha_{n'l'm';\omega_1}^A \Big)
        \\
        & ~~~~~~+ 
        \mathcal{H}^{(2)}_{(nlm);(n'l'm')} (X)
        \Big( \alpha_{nlm;\omega_1}^{A} ~\alpha_{n'l'm';-\omega_2}^{\bar{A}*} + \alpha_{nlm;\omega_2}^{\bar{A}} ~\alpha_{n'l'm';-\omega_1}^{A*} \Big)
                \\
        & ~~~~~~+ 
        \mathcal{H}^{(2)*}_{(nlm);(n'l'm')} (X)
        \Big( \alpha_{nlm;-\omega_2}^{\bar{A}*} ~\alpha_{n'l'm';-\omega_1}^{A} + \alpha_{nlm;-\omega_1}^{A*} ~\alpha_{n'l'm';-\omega_2}^{\bar{A}} \Big)
           \\
        & ~~~~~~+ 
        \mathcal{H}^{(1)*}_{(nlm);(n'l'm')} (X)
        \Big( \alpha_{nlm;-\omega_2}^{\bar{A}*} ~\alpha_{n'l'm';-\omega_1}^{A*} + \alpha_{nlm;-\omega_1}^{A*} ~\alpha_{n'l'm';-\omega_2}^{\bar{A}*} \Big)
       \Big)
\\
        \mathcal{M}^{(3)}_{\omega_1, \omega_2}(X)=&\sum_{nlm;n'l'm'}\Big(
        \mathcal{H}^{(1)}_{(nlm);(n'l'm')}  (X) ~\alpha_{nlm;\omega_1}^{\bar{A}} ~\alpha_{n'l'm';\omega_2}^A 
        +  \mathcal{H}^{(2)}_{(nlm);(n'l'm')}  (X) ~\alpha_{nlm;\omega_1}^{\bar{A}} ~\alpha_{n'l'm';\omega_2}^{\bar{A}*}
        \\
        &  ~~~+ \mathcal{H}^{(2)*}_{(nlm);(n'l'm')}  (X) ~\alpha_{nlm;-\omega_2}^{\bar{A}*} ~\alpha_{n'l'm';-\omega_1}^{\bar{A}} +  + \mathcal{H}^{(1)*}_{(nlm);(n'l'm')}  (X) ~\alpha_{nlm;-\omega_2}^{\bar{A}*} ~\alpha_{n'l'm';-\omega_1}^{\bar{A}*}\Big).
    \end{aligned}
\end{equation}

\subsection{The Final Formula}

In order to reconstruct the twirled Petz map formula, we need to put \eqref{leading} and \eqref{subleading} in the twirled Petz map \eqref{twiled}. Using the results in Sec. \ref{calculation}, and the fact that $ \tau_A = \tr_{\bar{A}}[P_{code}]$ commute with $ \tr _{\bar{A}}[P_{code} O P_{code}], \forall O \in\mathcal{L}(\mathcal{H}_{code})$ \cite{Bahiru:2022ukn}, we arrive at
\begin{equation}
    \begin{aligned}
        \Phi_A(X) = & \sum_ \omega \mathcal{F}^A_\omega(X) A_\omega 
        \\
        &~~+\frac{\lambda}{N} \sum_{\omega_1,\omega_2} \big( \mathcal{M}^{(1)}_{\omega_1, \omega_2} (X) A_{\omega_1} A_{\omega_2} +e^{\omega_2/2} \mathcal{M}^{(2)}_{\omega_1, \omega_2} (X) A_{\omega_1} A_{-\omega_2} + e^{(\omega_1 +\omega_2)/2} \mathcal{M}^{(3)}_{\omega_1, \omega_2} (X) A_{-\omega_1} A_{-\omega_2}\big)
        \\
        =&  \sum_ \omega \mathcal{F}^A_\omega(X) A_\omega 
        \\
        &~~+ \frac{\lambda}{N} \sum_{\omega_1,\omega_2} \big( 
      \mathcal{M}^{(1)}_{\omega_1, \omega_2}(X) +e^{- \omega_2/2} \mathcal{M}^{(2)}_{\omega_1, -\omega_2} (X) +  e^{-(\omega_1 +\omega_2)/2} \mathcal{M}^{(3)}_{-\omega_1, -\omega_2} (X) \big)~A_{\omega_1} A_{\omega_2}
    \end{aligned}
\end{equation}

To find the explicit form of the modular eigenfunction for the algebra corresponding to the entanglement wedge, we have \eqref{eigen}
\begin{equation}
    \mathcal{A}_{i,\nu} = \int ds ~ e^{is \nu}~ e^{is K_A}~a_i~ e^{-isK_A},~~~ \mathcal{A}_{\mathcal{E}_A} = \text{span}\{a_i\}.
\end{equation}
Therefore, our task is to find $\mathcal{A}_{\mathcal{E}_A}  $. Since at $ t=0$, we have $ \phi(t=0, r, \Omega) = \phi_0(t=0, r, \Omega) $, we reach \cite{faulkner2017bulk}
\begin{equation}
    \mathcal{A}_{\mathcal{E}_A}  = \text{span}\{O_0^c(t=0, x_A), \forall x_A \in A \}.
\end{equation}
Therefore, the eigenfunctions of the modular Hamiltonian given by two labels $\nu : ( \omega, x_A)$ as
\begin{equation}\label{eigenfun}
    A_\omega(x_A) = \int ds ~ e^{is\omega}~ e^{isK_A}~ O_0^c(t=0, x_A)~ e^{-isK_A}.
\end{equation}

As a result, the leading term can be rewritten as 
\begin{equation}
 \int_{-\infty}^{\infty} ds~ K^{(0)}_{Petz}(X|x_A,s)~ e^{isK_A}~ O_0^c(t=0, x_A)~ e^{-isK_A}
\end{equation}
where, by substituting $ O_0^c(t=0, x_A)=O^c(t=0, x_A)$, one can obtain that the $N^0 $ term as 
\begin{equation}
   \boxed{ \int_{-\infty}^{\infty} ds~ \int dx_A~K^{(0)}_{Petz}(X|x_A,s)~O^c(s,x_A)}
\end{equation}
where
\begin{equation}
   O^c(s,x_A)=  e^{isK_A}~ O^c(t=0, x_A)~ e^{-isK_A}
\end{equation}
where 
\begin{equation}
\begin{aligned}
        K^{(0)}_{Petz}(X|x_A,s) &=  \sum_\omega  ~ \mathcal{F}^A_{\omega, x_A}  (X) e^{is\omega}
        \\
        &= \sum _{nlm}  \sum_\omega~ e^{is\omega}~\Big( f_{nlm}(X) ~\alpha^A_{nlm;\omega} + f^*_{nlm}(X) ~\alpha^{A*}_{nlm;-\omega} \Big)
\end{aligned}
\end{equation}
which is the same result as the previous calculation.
By repeating the same substitution, one can find that the subleading term can be read off as 
\begin{equation}
    \boxed{\frac{\lambda}{N} \int_{-\infty}^{\infty} ds_1~ \int_{-\infty}^{\infty} ds_2 \int dx_A^1 dx_A^2 ~ K^{(1)}_{Petz}(X|x_A^1,x_A^2,s_1,s_2)~ O^c(s_1,x_A^1)~O^c(s_2,x_A^2)}
\end{equation}
where subleading kernel is
\begin{equation}
    K^{(1)}_{Petz}(X|x_A^1,x_A^2,s_1,s_2) = \sum_{\omega_1, \omega_2}~ e^{is\omega_1}~e^{is\omega_2}  \big( 
      \mathcal{M}^{(1)}_{\omega_1, \omega_2}(X) +e^{- \omega_2/2} \mathcal{M}^{(2)}_{\omega_1, -\omega_2} (X) +  e^{-(\omega_1 +\omega_2)/2} \mathcal{M}^{(3)}_{-\omega_1, -\omega_2} (X) \big).
\end{equation}
Finally, we have
\begin{equation}
    \begin{aligned}
        \Phi_A(X) =& \int_{-\infty}^{\infty} ds~ \int _{x_A \in  A} dx_A~K^{(0)}_{Petz}(X|x_A,s)~O^c(s,x_A)
        \\& ~~+ \frac{\lambda}{N} \int_{-\infty}^{\infty} ds_1~ \int_{-\infty}^{\infty} ds_2 \int _{x^1_A, x_A^2 \in  A}  dx_A^1 dx_A^2 ~ K^{(1)}_{Petz}(X|x_A^1,x_A^2,s_1,s_2)~ O^c(s_1,x_A^1)~O^c(s_2,x_A^2).
    \end{aligned}\label{codefinal}
\end{equation}
Like the leading order, the natural extension of the operator can be 
\begin{equation}
    \begin{aligned}
        \Phi_A(X) =& \int_{-\infty}^{\infty} ds~ \int _{x_A \in  A} dx_A~K^{(0)}_{Petz}(X|x_A,s)~O(s,x_A)
        \\& ~~+ \frac{\lambda}{N} \int_{-\infty}^{\infty} ds_1~ \int_{-\infty}^{\infty} ds_2 \int _{x^1_A, x_A^2 \in  A}  dx_A^1 dx_A^2 ~ K^{(1)}_{Petz}(X|x_A^1,x_A^2,s_1,s_2)~ O(s_1,x_A^1)~O(s_2,x_A^2).
    \end{aligned}
\end{equation}
However, we note that one can find other possible operators $\Phi_A(X) \in A$ in the CFT Hilbert space, but the action of all of them on the code subspace are the same as in \eqref{codefinal}.

In this section, we have completed the construction of entanglement wedge reconstruction up to first subleading order in 
$1/N$ by expressing the projected HKLL operators in the eigenbasis of the modular Hamiltonians associated with regions 
$A$ and 
$\bar{A}$. By exploiting the equivalence between boundary and bulk Bogoliubov transformations within the code subspace, we showed that the leading contribution to the reconstructed bulk field inside the entanglement wedge 
$a= \mathcal{E}_A$
 can be written entirely in terms of modular-flowed single-trace operators supported on 
$A$, reproducing the expected locality properties of entanglement wedge reconstruction. Beyond leading order, we derived an explicit bilocal correction governed by a subleading Petz kernel. The final result provides a concrete realization of quantum error correction in holography beyond the GFF limit, demonstrating how interacting bulk dynamics are encoded perturbatively in boundary modular data. This formulation sets the stage for addressing additional sources of corrections—such as gravitational dressing, crossed-product extensions, and genuinely dynamical geometries.

\section{Discussion}

In this work, we have provided a systematic derivation of the entanglement wedge reconstruction (EWR) for interacting bulk fields using the twirled Petz map, including the first-order 
$1/N$ corrections arising from cubic interactions in the bulk. Our analysis highlights several important features and implications.

Our derivation is general and applies to semiclassical AdS geometries without horizons, where the bulk spacetime remains approximately classical and time-translation invariant. In these backgrounds, the interacting bulk fields can be expanded in terms of mode operators that coincide with the free field modes at a reference time 
$t=t_0$. This identification allows us to perturbatively map interacting fields onto the generalized free field (GFF) code subspace, providing a controlled framework to compute both the leading and subleading contributions in 
$1/N$. As a result, by using the twirled Petz map, the fields inside the entanglement wedge  can be expressed entirely in terms of boundary modular operators and modular-flowed local operators of the corresponding region $A$, making explicit the role of the algebra of boundary operators in encoding the bulk region.

This construction also shows that the entanglement wedge can still be reconstructed perturbatively in interacting theories, with corrections systematically encoded in the subleading kernels 
$ \mathcal{M}_{\omega_1,\omega_2}^{(i)}(X)$ of the twirled Petz map. In particular, the leading 
$N^0$ term reproduces the modular flow reconstruction in \cite{faulkner2017bulk}, while the 
$1/N$ terms encode nontrivial operator mixing due to interactions and backreaction.
While we focused on cubic bulk interactions, the method generalizes to higher-order interactions and can systematically capture the full 
$1/N$ expansion of the twirled Petz map.

One can extend this analysis to geometries with horizons, which will allow us to study how interactions and modular flow modify the entanglement wedge behind the horizon, including implications for the information paradox. In order to explicitly do the reconstruction, one can use the notion of the algebraic Petz map discussed in \cite{Vardian:2023fce}.

Another important source of corrections arises from gravitational dressing. While our analysis has been carried out in a perturbative regime with small backreaction around a fixed semiclassical background, diffeomorphism invariance implies that strictly local bulk operators do not exist as gauge-invariant observables and must instead be defined relationally with respect to the boundary. Such gravitational dressing modifies the action of the modular Hamiltonian on bulk operators and, in principle, leads to additional contributions to the reconstruction kernels. Within our perturbative framework, these effects are expected to appear as higher-order corrections in the 
$1/N$ expansion of the twirled Petz map. In particular, for geometries dual to high-energy semiclassical states, one may employ the state-dependent dressing proposal of \cite{Bahiru:2022oas, Bahiru:2023zlc, Geng:2024dbl, Geng:2025rov, Geng:2025gns, Geng:2025bcb, Geng:2025byh, Geng:2025gqu, Geng:2026asi, elsayed2026observablealgebrajtdesitter}, which preserves approximate bulk locality while providing a well-defined mapping between bulk and boundary operators. This construction allows for entanglement wedge reconstruction in region 
$A$ in a manner consistent with relational locality, ensuring that the reconstructed operators remain approximately local within the entanglement wedge despite the presence of gravitational constraints. 

It would be interesting to relate the reconstruction error captured explicitly by the twirled Petz channel to the exponentially suppressed violations of exact EWR identified in \cite{Witten:2026gze}, and in particular to understand how the subleading Petz kernel is related to the state dependence of the effective area function.

Overall, our results demonstrate that entanglement wedge reconstruction in interacting theories is not only possible but can be formulated explicitly in terms of modular-flowed operators and perturbative kernels, providing a robust framework for connecting bulk interactions to boundary operator dynamics.

\section*{Acknowledgments}
We thank P. Hayden and K. Papadodimas for the valuable discussion and communication.
Authors also acknowledge the use of ChatGPT (OpenAI) as an editorial assistant in improving the clarity and presentation of the manuscript.

\nocite{*}

\bibliographystyle{ieeetr}
\bibliography{refs}

\end{document}